\documentclass[aps,prx,reprint,superscriptaddress]{revtex4-1}

\usepackage[utf8]{inputenc}

\usepackage{graphicx}
\usepackage{bm}
\usepackage{amsmath}
\usepackage{color,soul}
\soulregister\cite7
\soulregister\ref7
\usepackage[colorlinks,bookmarks=false,citecolor=blue,linkcolor=red,urlcolor=blue]{hyperref}

\definecolor{sectionnamecolor}{rgb}{0,0,0.5}
\definecolor{Juditcolor}{rgb}{1,0,0.5}
\definecolor{KPnotecolor}{rgb}{0.6,0.,0.4}

\newcommand{\sectionname}[1]{}

\begin{document}

%\title{Two magnon excitations of the multiferroic Sr$_2$CoGe$_2$O$_7$ in high magnetic fields.}
\title{Direct observation of spin-quadrupolar excitations in Sr$_2$CoGe$_2$O$_7$ by high field ESR}

\author{Mitsuru Akaki}
\email[]{akaki@ahmf.sci.osaka-u.ac.jp}
\affiliation{Center for Advanced High Magnetic Field Science, Graduate School of Science, Osaka University, Toyonaka, Osaka 560-0043, Japan}

\author{Daichi Yoshizawa}
\affiliation{Center for Advanced High Magnetic Field Science, Graduate School of Science, Osaka University, Toyonaka, Osaka 560-0043, Japan}

\author{Akira Okutani}
\affiliation{Center for Advanced High Magnetic Field Science, Graduate School of Science, Osaka University, Toyonaka, Osaka 560-0043, Japan}

\author{Takanori Kida}
\affiliation{Center for Advanced High Magnetic Field Science, Graduate School of Science, Osaka University, Toyonaka, Osaka 560-0043, Japan}

\author{Judit Romh\'{a}nyi}
\affiliation{Okinawa Institute of Science and Technology Graduate University, Onna-son, Okinawa 904-0395, Japan}

\author{Karlo Penc}
\affiliation{Institute for Solid State Physics and Optics, Wigner Research Centre for Physics, Hungarian Academy of Sciences, H-1525 Budapest, P.O.B. 49, Hungary}
\affiliation{MTA-BME Lend\"{u}let Magneto-optical Spectroscopy Research Group, 1111 Budapest, Hungary}

\author{Masayuki Hagiwara}
\affiliation{Center for Advanced High Magnetic Field Science, Graduate School of Science, Osaka University, Toyonaka, Osaka 560-0043, Japan}

\date{\today}

\begin{abstract}
Exotic spin-multipolar ordering in spin transition metal insulators has so far eluded unambiguous experimental observation. 
A less studied, but perhaps more feasible fingerprint of multipole character emerges in the excitation spectrum in the form of quadrupolar transitions. 
Such multipolar excitations are desirable as they can be manipulated with the use of light or electric field and can be captured by means of conventional experimental techniques. 
Here we study single crystals of multiferroic Sr$_2$CoGe$_2$O$_7$, and observe a two-magnon spin excitation appearing above the saturation magnetic field in electron spin resonance (ESR) spectra. 
Our analysis of the selection rules reveals that this spin excitation mode does not couple to the magnetic component of the light, but it is excited by the electric field only, in full agreement with the theoretical calculations. 
Due to the nearly isotropic nature of Sr$_2$CoGe$_2$O$_7$, we identify this excitation as a purely spin-quadrupolar two-magnon mode.  

\end{abstract}

\pacs{76.60.Gv, 75.85.+t, 76.50.+g, 75.50.Ee}

\maketitle

%%%%%%%%%%%%%%%%%%%%%%%%%%%%%%%%%%%%%%%%%%%%%%%%
\section{Introduction}
%%%%%%%%%%%%%%%%%%%%%%%%%%%%%%%%%%%%%%%%%%%%%%%%

The absence of spatial inversion and time reversal symmetries may lead to the magneto-electric effect, where the magnetization and the electric polarization of a material are coupled, allowing for the mutual control of magnetization by electric and polarization by magnetic fields, providing new multiferroic materials for future technologies ~\cite{ME1,ME2,ME3,KimuraN}. 
Well known examples are vector spin chirality and exchange striction driven electricity ~\cite{Katsura, Sergienko, Jia}, both involving more than one spin. In the vector spin chirality mechanism, the non-collinearity of the neighboring magnetic moments may induce electric polarization, while in the exchange striction case the charged magnetic ions move to optimize the Heisenberg exchange energy between the neighbours with parallel and antiparallel magnetic moments.

%------      FIG 1    ---------------------------------------------------------------------------------------
\begin{figure}[b]
\begin{center}
\includegraphics[width=0.8\columnwidth]{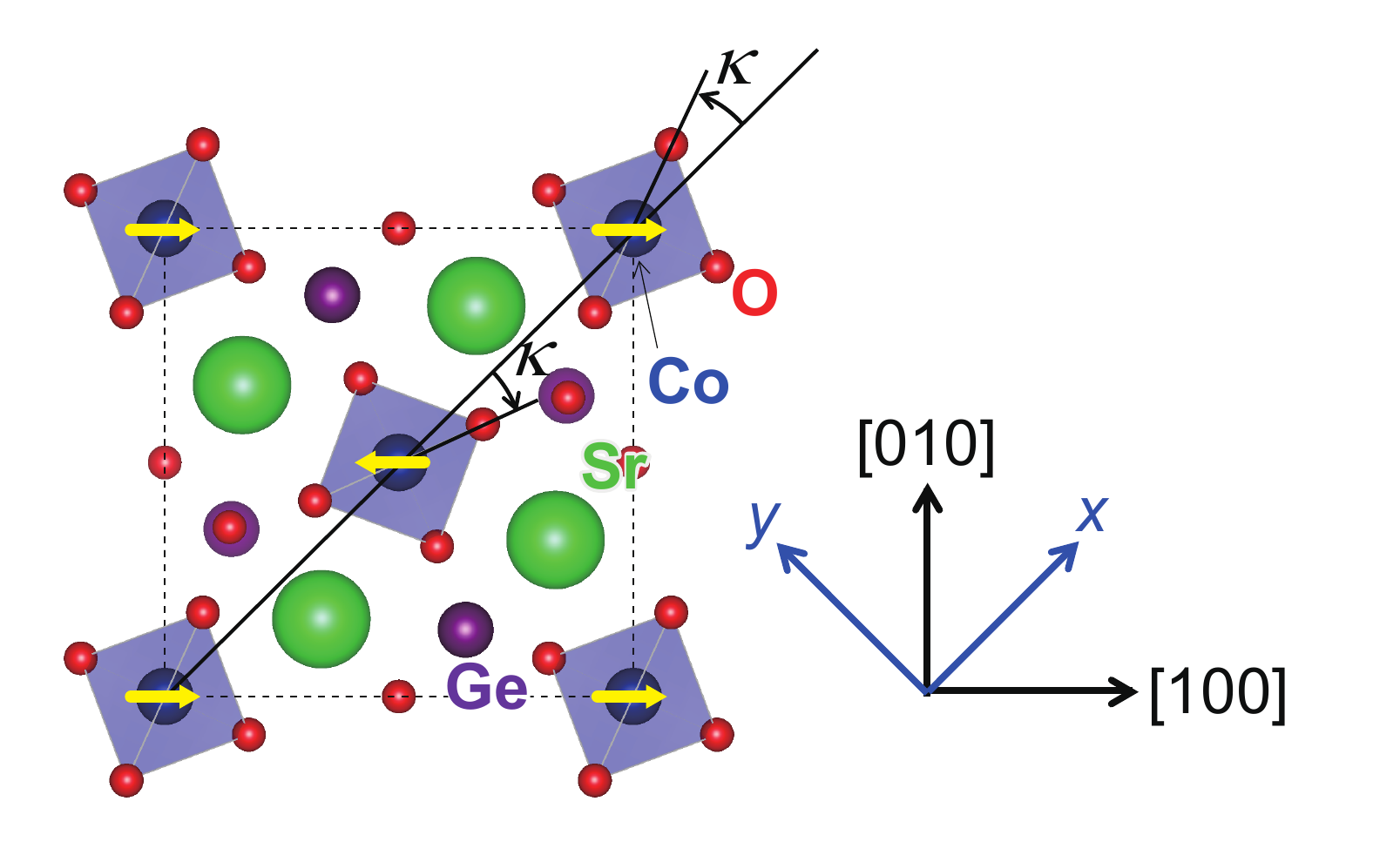}
\vspace{-9pt}
\caption{\label{fig:structure} The schematic crystal and magnetic structures of SCGO projected onto the $ab$--plane. The blue spheres represent the magnetic Co$^{2+}$ ions with $S=3/2$ surrounded by four O$^{2-}$ ions in a tetrahedral environment. The two tetrahedra in the unit cell (dashed square) are rotated alternatively by an angle $\kappa$. The yellow arrows show the spin directions in the ordered phase in the absence of external magnetic field. On the right hand side we show the crystallographic coordinate system with [100] and [010] axes, as well as the $x$ and $y$ coordinates, following the convention of Ref.~[\onlinecite{MiyaharaJPSJ}]  }
\end{center}
\end{figure}

Remarkably, in \aa kermanites -- where CoO$_4$ tetrahedra form diagonal square lattices, alternating with intervening layers of alkaline earth metal ions along the $c$-axis  [Fig.~\ref{fig:structure}] -- the electric polarization is not induced by correlations between neighbouring spins, but is also present in the paramagnetic phase~\cite{AkakiPRB}. 
These compounds represent an exceptional family of magneto-electric materials, in which the finite (on-site) polarization emerges on the account of relativistic metal-ligand hybridization~\cite{Arimapd}.
For this mechanism at least two conditions have to meet: The transition metal ion has to carry a spin larger than 1/2 and the inversion symmetry on this site needs to be broken.  
The spin-3/2 magnetic moment of Co$^{2+}$ allows for spin-quadrupole operators, which are time reversal invariant quadratic expressions of local spin. 
The lack of inversion symmetry at the tetrahedrally coordinated Co sites permits the magneto-electric effect, with on-site polarization proportional to the quadrupole operators already at a level of a single ion. 
The metal-ligand hybridization is believed to act in Ba$_2$CoGe$_2$O$_7$~\cite{YiAPL,MurakawaPRL}, Sr$_2$CoSi$_2$O$_7$~\cite{AkakiPRB}, and Ca$_2$CoSi$_2$O$_7$~\cite{AkakiAPL,AkakiJPSJ}. 
This mechanism is considered to be responsible for the dynamical properties of these materials, as it couples the spin degrees of freedom to the oscillating electric field component of the light. 

\AA kermanites are usually characterized by large easy-plane magnetic anisotropy and  small exchange interaction. As a consequence the spin dipole and quadrupole degrees of freedom become mixed and it is challenging to untangle the different types of multipolar fluctuations in the excitation spectra. 
This is well exemplified in the THz absorption spectra by the observation of the `electromagnon' in Ba$_2$CoGe$_2$O$_7$, an electrically active magnetic excitation having both dipolar and quadrupolar characters~\cite{Kezsmarki2011,MiyaharaJPSJ,KezsmarkiNatCom2014}.

Here we report on the properties of Sr$_2$CoGe$_2$O$_7$ (SCGO), a member of the \aa kermanite family with almost isotropic magnetization properties. 
When the magnetic anisotropies are small or absent, mixing of the spin dipole and quadrupole degrees of freedom is suppressed, reflecting the higher symmetry of the system. 
Using electron spin resonance (ESR) technique supported by theoretical calculations, we show that, due to its isotropic nature, SCGO  exhibits a purely quadrupolar two-magnon mode in high magnetic fields. 
Measuring in different geometries for both Faraday and Voigt configurations, we find that this magnetically inactive excitation can only be excited by specific components of the oscillating electromagnetic field, in full agreement with the predictions of the relativistic metal-ligand hybridization. 

The article is structured as follows: In Sec.~\ref{sec:sample} we give details about the sample and experimental methods. In Sec.~\ref{sec:Ham} we introduce the spin quadrupoles and the magnetoelectric coupling in \aa kermanites, and we present the model Hamiltonian for the SCGO.  The static properties -- magnetization and electric polarization  -- of the SCGO are discussed in Sec.~\ref{sec:Statics}
, while the dynamical properties are considered in Sec.~\ref{sec:ESR_dynamics}, both experimentally and theoretically. Evidence for the quadrupolar nature of the two-magnon excitations is presented in Sec.~\ref{sec:selection_rules}, where the selection rules are examined. Finally, we conclude with a summary of our results in Sec.~\ref{sec:conlusion}.

%%%%%%%%%%%%%%%%%%%%%%%%%%%%%%%%%%%%%%%%%%%%%%
\section{Sample characterization, experimental details}
%%%%%%%%%%%%%%%%%%%%%%%%%%%%%%%%%%%%%%%%%%%%%%%%%
\label{sec:sample}

We grew single crystalline samples of SCGO using the floating zone method. 
Room temperature X-ray diffraction measurements confirmed the tetragonal $P\overline{4}2_1m$ structure with no impurity phases. All the samples used for the experiments were cut along the crystallographic principal axes to make samples with plate-like shapes after checking by the X-ray back-reflection Laue technique.

The magnetization was measured in static magnetic fields of up to 7~T using a commercial SQUID magnetometer and by the induction method, using a coaxial pick-up coil, in pulsed fields of up to 55~T, with a pulse duration of 7~millisecond.

The electric polarization induced by magnetic fields was obtained by integrating the polarization current as a function of time. Since in \aa kermanite materials  electric polarization emerges even without applying poling electric fields, which is a unique feature of their multiferroicity~\cite{AkakiAPL,AkakiPRB,MurakawaPRL,AkakiJPSJ}, we measured the polarization current without poling electric fields.

Low-field ESR spectra up to 14~T at $1.5-1.6$~K and for frequencies below 500~GHz were taken using a home-made transmission ESR cryostat in a superconducting magnet.

\begin{figure}

\begin{center}
\includegraphics[width=0.9\columnwidth]{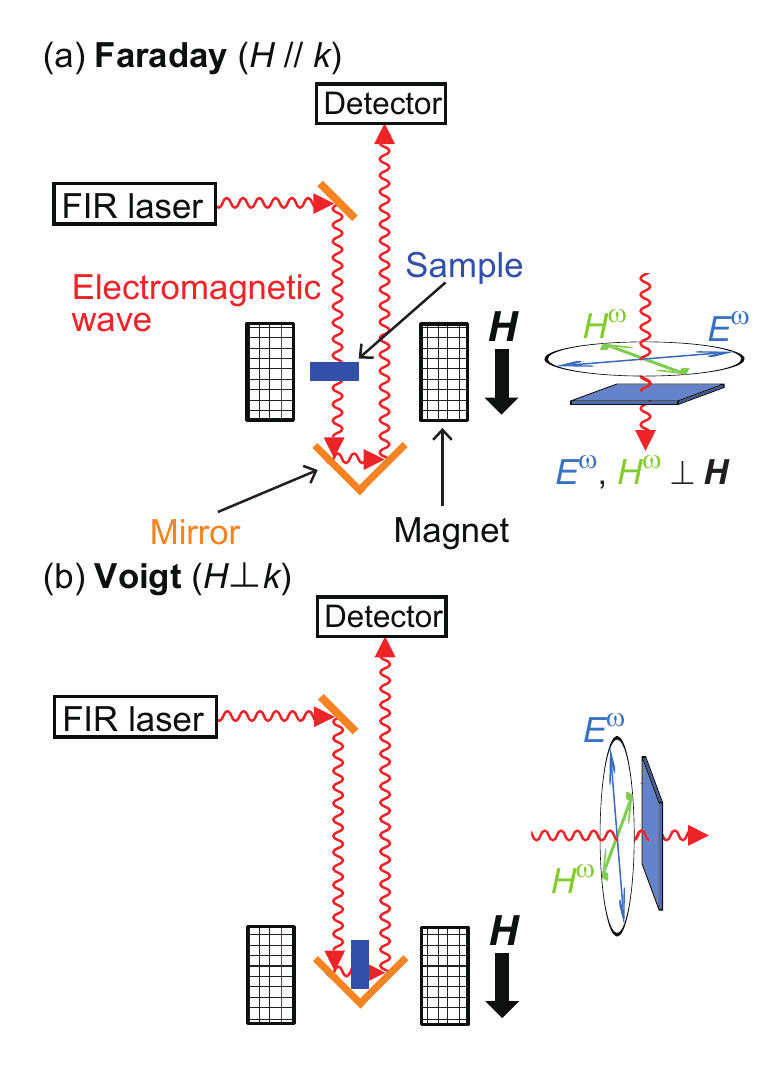}
\vspace{-9pt}
\caption{\label{fig:esr_apparatus} Schematic experimental setup for the high-field ESR measurements, with different light configurations. $E^{\omega}$ ($H^{\omega}$) shows one of the electric (magnetic) component of the unpolarized light.}
\end{center}
\end{figure}

High-field ESR measurements at 1.4~K in pulsed magnetic fields of up to 55~T were conducted by utilizing a far-infrared laser and Gunn oscillators (75, 90, 95, 110, and 130 GHz)  coupled with a frequency doubler to generate sub-millimeter and millimeter waves. We used an InSb bolometer as a detector. All the experiments were carried out using unpolarized light.
Figure.~\ref{fig:esr_apparatus} shows the schematic experimental setup of the ESR spectrometer in pulsed magnetic fields.

%-----------------------------------------------------------------------------------------------------------------------
%------      FIG 3    ---------------------------------------------------------------------------------------
%-----------------------------------------------------------------------------------------------------------------------

%%%%%%%%%%%%%%%%%%%%%%%%%%%%%%%%%%%%%%%%%%%%%%%%
\section{Magnetoelectric coupling and the Hamiltonian in \aa kermanites}
%%%%%%%%%%%%%%%%%%%%%%%%%%%%%%%%%%%%%%%%%%%%%%%%
\label{sec:Ham}

\subsection{Magnetoelectric coupling}  
\label{sec:Ham_ME}

In \aa kermanites the spin vector chirality $\mathbf{\hat{S}}_i\times\mathbf{\hat{S}}_j$ is negligible and the exchange interaction is uniform for each bond disabling the spin current or exchange striction as the origin of spin induced polarization \cite{MurakawaPRL}. Instead, spin-dependent metal-ligand hybridization has been put forward as the source of finite polarization, induced by onsite second order spin-terms, in contrast to the aforementioned concepts which involve two neighboring spins. 

Since the polarization vector is 
%even under time reversal and 
odd and the magnetic moment (spin) 
%is odd under time reversal and 
is even under spatial inversion, the spins can induce polarization only if the inversion symmetry is absent. For an ion in such a non-centrosymmetric site, the general form for the $\alpha={x,y,z}$ components of the electric polarization at site ${\bf r}$ is
\begin{equation}
 P^\alpha_{\bf r} = \sum_{\beta,\gamma\in{x,y,z}} c^\alpha_{\beta\gamma} \hat{S}^\beta_{\bf r} \hat{S}^\gamma_{\bf r}  \;,
 \label{eq:PSS}
\end{equation}
where the $\hat{S}^\beta_{\bf r}$ and $\hat{S}^\gamma_{\bf r}$ are the spin components at the same site ${\bf r}$, and  $c^\alpha_{\beta\gamma}=c^\alpha_{\gamma\beta}$ with $\sum_\beta c^\alpha_{\beta\beta} = 0$ is a traceless  $3\times 3\times 3$ tensor symmetric in the lower two indices, effectively coupling the polarization to a time reversal invariant spin-quadrupole operators\cite{Inttab}. 
The symmetry properties of the crystal reflected by the local environment of the magnetic ions determine the explicit form of tensor $c^\alpha_{\beta\gamma}$ resulting in the following spin-induced polarization characteristic for the \aa kermanite family~\cite{MurakawaPRL,RomPRB,SodaPRL}:
\begin{subequations}
\begin{align}
\begin{pmatrix}
{P}^{x}_{j}\\
{P}^{y}_{j}
\end{pmatrix}
&\propto&
\begin{pmatrix}
 -\cos2\kappa  & - (-1)^j \sin2\kappa\\
- (-1)^j \sin2\kappa & \cos2\kappa  
\end{pmatrix} \cdot
\begin{pmatrix}
 \hat{Q}^{2xz}_j \\
 \hat{Q}^{2yz}_j 
\end{pmatrix} \;,
\label{eq:Pxy}
\end{align}
transforming as a two--dimensional irreducible representation of the point group, and
\begin{align}
P^{z}_{j} = -W_z[ 
\cos 2\kappa \,\hat{Q}^{x^2-y^2}_j
+ (-1)^j \sin2\kappa\, \, \hat{Q}^{2xy}_j] \;,
\label{eq:Pz}
\end{align}\label{PolOp}
\end{subequations}
a one--dimensional irreducible representation.
The $W_z$ is a coupling constant.
The factor $(-1)^j$  accounts for the alternation of the angle $\kappa\approx\pm 20.5^\circ$ on the two sublattices and $\kappa$ measures the rotation of the oxygen tetrahedra around the cobalt ions with respect to the [110] direction, as indicated in Fig.~\ref{fig:structure}.

The $\hat{Q}$ operators represent a symmetric combination of the spin operators
\begin{subequations}
\begin{align}
\hat{Q}^{2\alpha\beta}_j &= \hat{S}^\alpha_{j}\hat{S}^\beta_{j}+\hat{S}^\beta_{j}\hat{S}^\alpha_{j} \,, \\
\hat{Q}^{x^2-y^2}_j &= (\hat{S}^x_{j})^2 - (\hat{S}^y_{j})^2 \,, \\
\hat{Q}^{3z^2-r^2}_j &=   \frac{1}{\sqrt{3}} \left[ 3(S^z_j)^2 - \mathbf{S}_j\cdot \mathbf{S}_j  \right]\,. 
\end{align}
\end{subequations}
where the last one does not appear in the expressions for the polarization, Eqs.~(\ref{PolOp}). We include it, since they form the five spin-quadrupole operators \cite{quadrupolreview}, satisfying the 
\begin{equation}
  \sum_{\mu=x,y,z} \left[\left[\hat{Q}^{\eta}_j, \hat{S}_j^\mu \right],\hat{S}_j^\mu \right] = k (k+1) \hat{Q}^{\eta}_j 
\end{equation} 
property of a rank-$k$ tensor operator with $k=2$, where $\eta \in \{ 2yz, 2xz , 2xy, x^2-y^2, 3z^2-r^2\}$. This classification is valid for the systems with the $O(3)$ symmetry. The external magnetic field, however, lowers the O(3) symmetry of the space to the $C_{\infty h} +\Theta \sigma_v C_{\infty h}$ magnetic point group, where the axis of the $C_{\infty h}$ axial group is parallel to the magnetic field, $\Theta$ is the time reversal operator, and $\sigma_v$ is a reflection to a plane that includes the axis of the magnetic field \cite{shannonmotome2010}. When restricted to spins, this symmetry group is equivalent to the SO(2). In this case it is convenient to group the spin operators according to their transformation under the remaining SO(2) rotations about the direction of magnetic field --- conventionally the $z$ axis, as shown in table~\ref{tab:spinirreps}, but the coordinates shall be rotated according to the actual direction of the field.
Each of the separate irreps transforms like $e^{in\phi}$ --- where $\phi$ is the polar angle in the $xy$ plane 
perpendicular to the magnetic field. Actually the $S=3/2$ spin is large enough to support rank-3 tensor operators (octupoles), constructed by symmetric combinations of three spin operators, but we will neglect them here.
All these operators commute with an isotropic spin-Hamiltonian, and since they belong to different irreducible representations, they excite different modes which do not mix.

\begin{table} 
\caption{\label{tab:spinirreps} Classification of tensor operators according to rotational symmetry about a $z$ axis defined 
by magnetic field, isomorphic to SO(2).
Each forms an irreducible representation transforming like $e^{i n \phi}$, where $n$ is an integer and $\phi$ is the polar angle in the $xy$ plane. These operators commute with an isotropic spin Hamiltonian.  
} 
\begin{ruledtabular} 
\begin{tabular}{lccr} 
ired. repr. & dipole & quadrupole & $\Delta S^z$ \\ 
\hline
$e^{2i\phi}$  &                      & $Q^{x^2-y^2}+ i Q^{xy}$ & $+2$ \\ 
$e^{i\phi}$   & $S^{x}+i S^{y}$ & $Q^{xz} + i Q^{yz}$      & $+1$  \\ 
1             & $S^z$                & $Q^{3z^2-r^2}$           &  0 \\ 
$e^{-i\phi}$  & $S^{x}-i S^{y}$ & $Q^{xz} - i Q^{yz}$      & $-1$   \\ 
$e^{-2i\phi}$ &                      & $Q^{x^2-y^2} - i Q^{xy}$ & $-2$ \\ 
\end{tabular} 
\end{ruledtabular} 
\end{table}

The spin-dipolar and spin-quadrupolar operators  in the $e^{-i\phi}$  irreducible representation are $\propto S^-$, and therefore they create a single magnon with $\Delta S^z = -1$. The spin-quadrupolar operators in the $e^{-2i\phi}$ irreducible representation are $\propto S^-S^-$ and excite two magnons with $\Delta S^z = -2$ on a single site, and we will be mostly interested in them in this paper.
We shall note that the possibility to create two magnons by quadrupolar operators in Eqs.~(\ref{PolOp}) is in contrast with the spin-current mechanism, where the magnetoelectric coupling involves a bilinear, but spin-dipole operator only (a tensor operator with $k=1$), restricting the number of created magnons to one. 

The metal-ligand hybridization mechanism has been proposed as the microscopic origin of the polarization\cite{Arimapd} (see also [\onlinecite{Jia}]) , leading to 
\begin{equation}
  \mathbf{P}_j \propto \sum_{i=1}^4 \mathbf{e}_{i,j} \left[3 (\mathbf{e}_{i,j} \cdot \mathbf{\hat{S}}_j)^2 - \mathbf{\hat{S}}_j^2 \right]\;,
\end{equation} 
where the sum is over the four oxygens surrounding the Co magnetic ion at site $j$ and $\mathbf{e}_{i,j}$ denotes the unit vector pointing from the central Co$^{2+}$ ion to the $i$-th O$^{2-}$  ion. Performing the summation, we arrive at Eqs.~(\ref{PolOp}).
The $\mathbf{P}_j$ as defined above is a traceless operator in the spin Hilbert space. 

\subsection{Model Hamiltonian} 

As indicated by magnetic studies~\cite{EndoIC} and neutron dispersion measurements~\cite{Zheludev2003,SodaPRL}, the cobalt planes are  weakly coupled in \aa kermanites. Therefore we consider the following 2D spin Hamiltonian as a minimal microscopic model for SCGO \cite{MiyaharaJPSJ,RomPRB},
\begin{align}
\mathcal{H} & = 
 J\sum_{(i,j)}\left( \hat{S}^x_i  \hat{S}^x_j+ \hat{S}^y_i  \hat{S}^y_j\right)
+J_z \sum_{(i,j)} \hat{S}^z_i  \hat{S}^z_j
\nonumber\\& \phantom{=}
 +  J_{pz} \sum_{\langle i,j \rangle} P_i^zP_j^z 
+  \Lambda \sum_{i} (\hat{S}_i^z)^2 
\nonumber\\& \phantom{=}
- \mu_{\rm B} \sum_{i} [g_{ab}(H_x \hat{S}_i^x + H_y \hat{S}_i^y) + g_c H_z \hat{S}_i^z]\;,
\nonumber\\& \phantom{=}
 - g_s \mu_{\rm B} \sum_{i} (-1)^i \left(H_x S^y_i - H_y S^x_i \right)  \;
\label{eq:Hamiltonian}
\end{align}
where $J$ and $J_z$ are the anisotropic exchange constants between the nearest neighbor spins, $ \Lambda >0$ is the single ion anisotropy constant of the form imposed by the tetragonal symmetry, and $J_{pz}$ is the antiferroelectric coupling constant. We neglect the Dzyaloshinskii-Moriya interaction which is rather small even in more anisotropic \aa kermanites, and show that both static and dynamical properties can be reproduced for isotropic exchange interaction $J=J_z$ and with small single-ion anisotropy $ \Lambda \sim J \sim 1 K$. The staggered off-diagonal component of the $g$-tensor $g_s$ will play a role when we discuss the selection rules for the magnetic transitions.
 
Without the $J_{pz}$ term the Hamiltonian represents an easy plane antiferromagnet with a Goldstone mode ($\Delta_1=0$).
A finite $J_{pz}$ was introduced for Ba$_2$CoGe$_2$O$_7$ to explain the vanishing of electric polarization in zero field~\cite{RomPRB} and the absence of Goldstone modes in the neutron scattering study~\cite{SodaPRL}.
At zero magnetic field, the electric polarization is also zero in SCGO (Fig.~\ref{fig:staticpolarization}(b)), hence the ground state must be an antiferroelectric state, selecting [100] as the magnetic easy axis within the easy plane.

%%%%%%%%%%%%%%%%%%%%%%%%%%%%%%%%%%%%%%%%%%%%%%
\section{Static properties and magnetic anisotropy}
%%%%%%%%%%%%%%%%%%%%%%%%%%%%%%%%%%%%%%%%%%%%%%
\label{sec:Statics}

%------      FIG 3    ---------------------------------------------------------------------------------------
\begin{figure}
\begin{center}
\includegraphics[width=0.8\columnwidth]{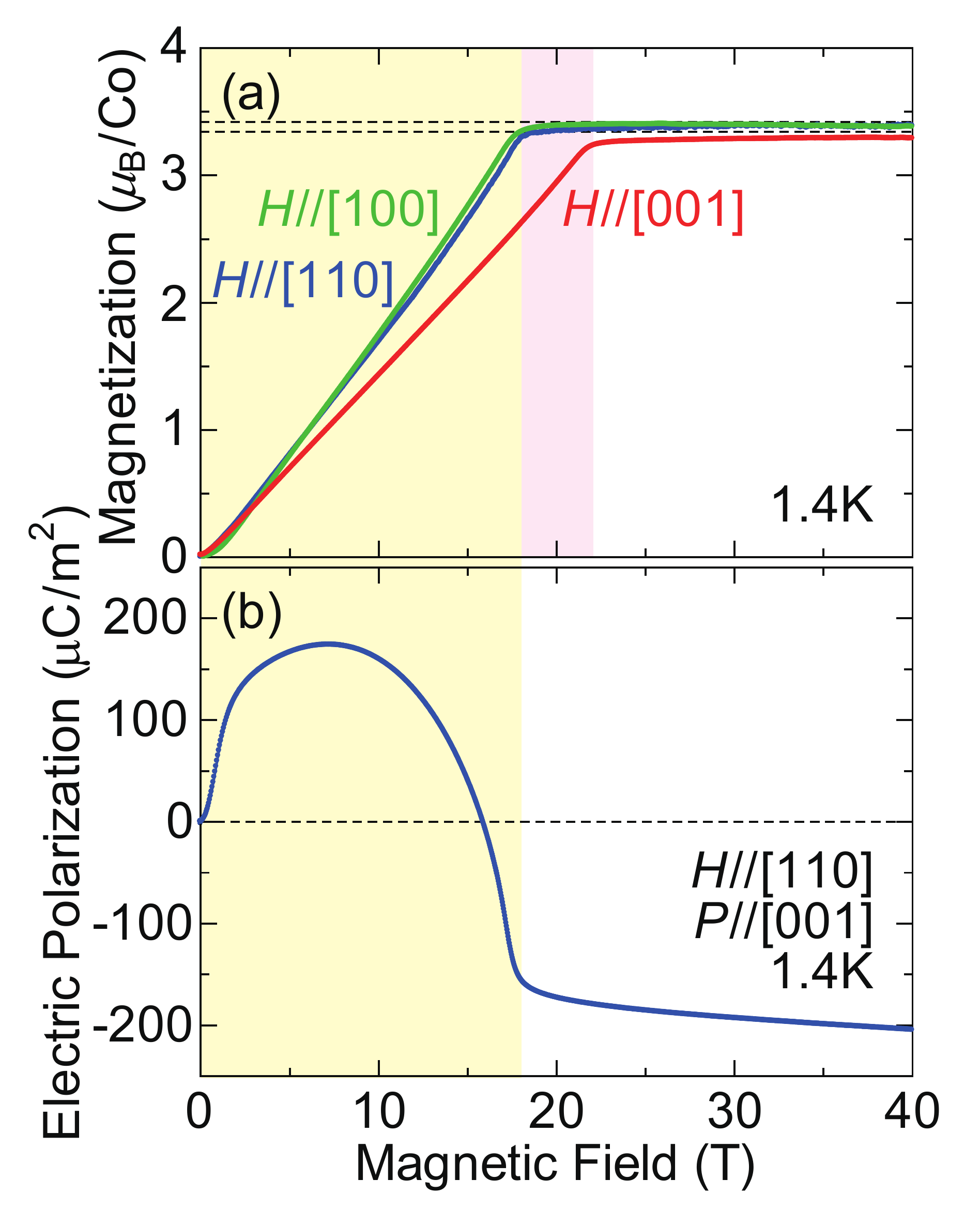}
\vspace{-9pt}
\caption{\label{fig:staticpolarization} (a) Magnetic field dependence of magnetization  in SCGO at 1.4~K\@.
Dashed lines indicate the saturation magnetization values using $g_{ab}=2.28$ and $g_c=2.23$, obtained from the ESR measurements [see Eqs.(\ref{eq:gD1[110]fit}) and (\ref{eq:gD1[001]fit})], for the in-plane and out-of-plane components, respectively. (b) Magnetic field dependence of the out-of-plane component of induced electric polarization ($P^z$) at 1.4~K\@ in the case of $\mathbf{H}\|[110]$. The in-plane polarization vanishes for this direction of the field. Above saturation, both the polarization and the magnetization become flat. Yellow and pink shaded areas mark the phases below saturation for in-plane and out-of-plane magnetic fields, respectively.}
\end{center}
\end{figure}

In order to obtain information about the magnetic anisotropies, we measured the magnetization and the induced electric polarization as function of external magnetic field. The results of our measurements, taken  at 1.4~K, well below the N\'eel temperature of $6.5$~K~\cite{EndoIC}, are displayed in Fig.~\ref{fig:staticpolarization}.
%

%-----------------------------------------------------------------------------------------------------------------------
\subsection{Magnetization}
\label{sec:magnetization}
%-----------------------------------------------------------------------------------------------------------------------

The magnetization in the cobalt plane, (001), rises almost linearly with magnetic field, reaching saturation at $3.42~\mu_{\rm B}/{\rm Co}$ just above 18~T\@. Along the perpendicular [001] direction, the corresponding values are $3.3~\mu_{\rm B}/{\rm Co}$ and 21.6~T, providing  $g_{ab}\approx2.3$ and $g_c \approx 2.2$  (we will get more precise values from fitting the ESR spectra in Sec.~\ref{sec:ESR_dynamics_fit}).
The magnetization curves are nearly isotropic, and the difference in the saturation fields is mainly explained by the $g$-tensor anisotropy.
In other \aa kermanite compounds, like Sr$_2$CoSi$_2$O$_7$~\cite{AkakiPRB} and Ba$_2$CoGe$_2$O$_7$~\cite{HutanuPRB}, the slope of in-plane magnetization is almost twice the slope of the  magnetization measured perpendicular to the cobalt plane. This large difference signals the presence of large easy-plane anisotropy. Here, the discrepancy between the slopes of in-plane and out-of-plane magnetization is small, and can be accounted for by a smaller $g$-tensor anisotropy and a smaller easy-plane anisotropy.  

To quantify how different the magnetic anisotropies are in SCGO compared to the sister materials, let us take a look at the saturation fields. As shown in the 
Appendix~\ref{sec:in_plane} and~\ref{sec:outof_plane}, the saturation fields from a mean field approach can be expressed as $\mu_{\rm B} g_{ab}H^{\text{sat}}_{xy}=12 J$ and $\mu_{\rm B} g_{c}H^{\text sat}_{z}=6 J+6J_z+2\Lambda$ for the in-plane and out-of-plane magnetic field directions. We can easily utilize this discrepancy in the saturation fields to get an estimate for the easy-plane and exchange anisotropies:
\begin{equation}
 6(J_z-J)+2 \Lambda = \mu_{\rm B} g_c H_z^{\text{sat}} - \mu_{\rm B} g_{ab} H_{xy}^{\text{sat}} .
\end{equation}
 Assuming  a negligibly small exchange anisotropy (which will be justified when we analyze the ESR data in Sec.~\ref{sec:ESR_dynamics_fit}), we get $ \Lambda \approx 2.2$ K for SCGO, while for Ba$_2$CoGe$_2$O$_7$ for example, the difference in saturation fields in the easy plane, (001), and hard axis direction, [001], is about $20$~T, resulting in an order of magnitude larger anisotropy of $13$~K~\cite{PencPRL}.
 
The ratio %$ \Lambda /J = 1.01$
$ \Lambda /J$ is closely related to the size of the ordered moment in the variational (mean field) approximation, larger anisotropy results in greater spin reduction: 
 \begin{equation}
 \langle S \rangle = \frac{3}{2} \left[1 - \left(\frac{\Lambda}{12 J}\right)^2 \right]
 \label{eq:spin_lenght}
 \end{equation}
 
As in SCGO the $ \Lambda /J $ is an order of magnitude smaller than in the other \aa kermanites~\cite{MiyaharaJPSJ,PencPRL}, the projection of the spin length along the field is nearly the maximal 3/2, and  the dipole and quadrupole characters of the excited modes are well separated. 
 
 The magnetization shown here is the measured one, the van-Vleck term is not subtracted. The magnetization above the saturation field is very flat, the van-Vleck paramagnetic susceptibility is vanishingly small. This is in line with the small anisotropies, as the level splitting of $t_2$ orbital is small and the CoO$_{4}$ tetrahedra are weakly distorted~\cite{YamauchiPRB}.

%-----------------------------------------------------------------------------------------------------------------------
\subsection{Electric polarization}
%-----------------------------------------------------------------------------------------------------------------------

The behavior of electric polarization as a function of magnetic field is nearly the same as in other \aa kermanites~\cite{AkakiPRB,MurakawaPRL}. 
However, subtle deviations provide additional proof of a smaller easy-plane anisotropy. When the field is applied in the $ab$ cobalt plane, the polarization has only out-of-plane component, {\it i.e.} only $P^z$ is finite. The amplitude of the $P^z$ changes as the magnetic field is rotated within the $ab$ plane: it's absolute value is maximal for the $H||[110]$ (and changes sign for  $H||[1\bar10]$), shown in Fig.~\ref{fig:staticpolarization}(b) as a function of magnetic field, and $P^z$ vanishes when the field is along the [100] or [010] directions. This agrees with the behavior seen in Sr$_2$CoSi$_2$O$_7$\cite{AkakiPRB} and and Ba$_2$CoGe$_2$O$_7$~\cite{MurakawaPRL}, and is consistent with Eq.~(\ref{eq:Pz}).
The inflection point of the polarization curve at $18$~T signals the transition to the saturated phase. Before reaching the transition point the spins are turning within the cobalt-plane towards the field direction, in this case towards [110]. Due to the easy plane anisotropy, they are not fully grown 3/2 spins, but somewhat shorter, see Eq.~(\ref{eq:spin_lenght}). As they turn towards the field, the polarization changes, reaching its extrema when the spins are parallel to one of the tetrahedron edges. Note that if the tetrahedra were not rotated, {\it i.e.} if $\kappa$ were zero, the polarization would have its extrema when the spins are aligned with [110] and [$\overline{1}$10].
At the transition field, the spins do not rotate any further, but the magnetic field, now strong enough to compete with the anisotropy, stretches them to asymptotically reach their maximal 3/2 value. This behavior can be nicely seen in magnetization measurements in the other members of the \aa kermanite family, as the magnetization after saturation is not a completely field independent constant, but further increases  towards the full saturation value~\cite{HutanuPRB,AkakiPRB}. This is also apparent in the polarization curve which decreases slowly after the transition instead of becoming flat~\cite{AkakiPRB}. More prominent changes of these observables after the saturation indicate that the spins are further away from being fully grown, and consequently that the material has larger single-ion anisotropy. 
In the case of SCGO, this decay in $P^z$ and the climb in the magnetization is significantly smaller than those in Sr$_2$CoSi$_2$O$_7$ and Ba$_2$CoGe$_2$O$_7$, further evidencing a smaller anisotropy.

To conclude the experimental observations of the static properties, the easy-plane single ion anisotropy in SCGO supports a planar antiferromagnetic configuration of cobalt spins in agreement with neutron powder diffraction measurements~\cite{EndoIC}, similarly to its sister compounds. However, both magnetization and polarization data indicate that the anisotropy is about an order of magnitude smaller than in the previously studied compounds, and the spin lengths in the ground state are much closer to the isotropic 3/2 value.

%------      FIG 4    ---------------------------------------------------------------------------------------
\begin{figure}[ht!]
\begin{center}
\includegraphics[width=0.8\columnwidth]{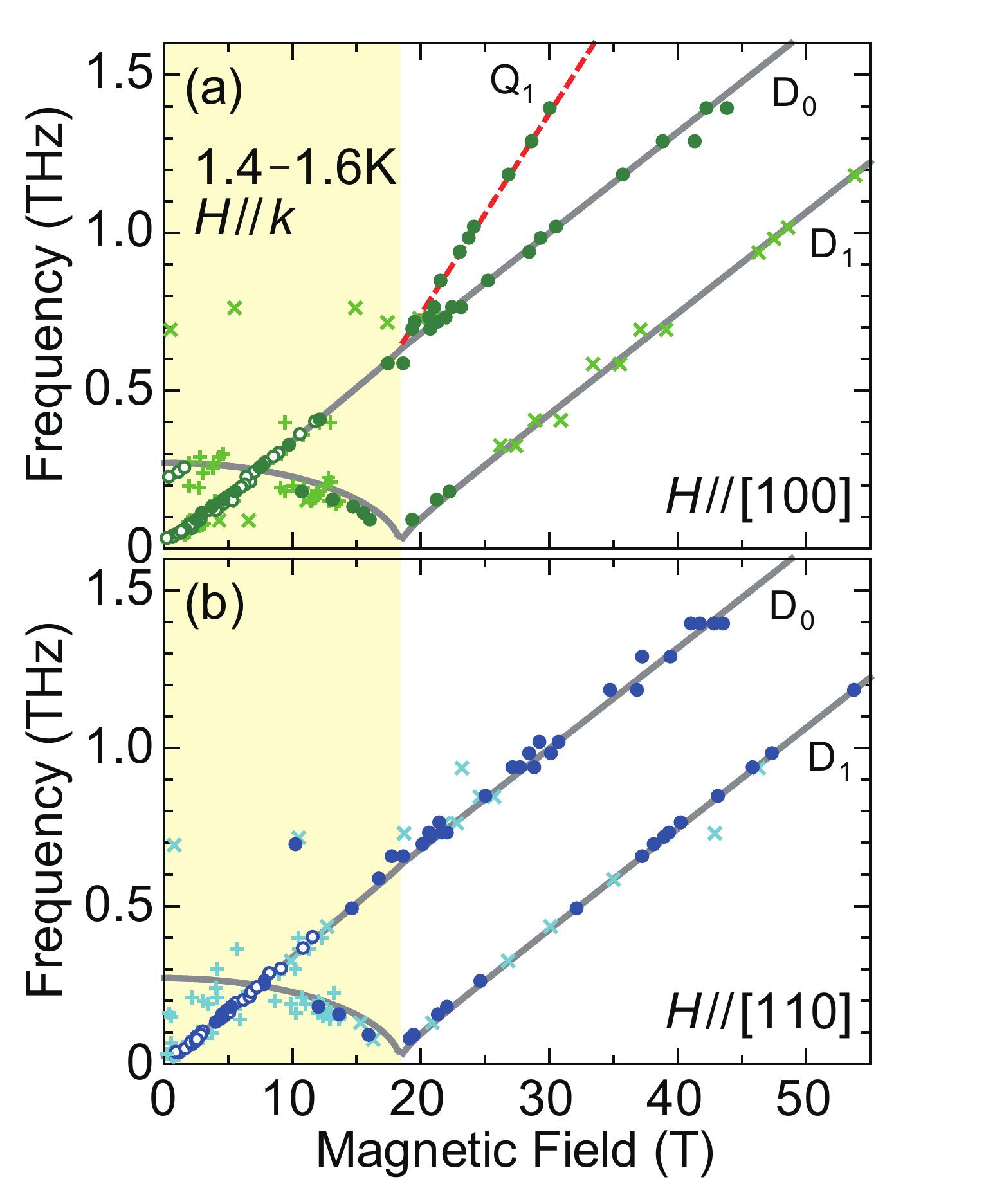}
\vspace{-9pt}
\caption{\label{fig:ESR_100_110}  Frequency-field diagrams of the ESR resonance fields of Sr$_2$CoGe$_2$O$_7$ for magnetic fields parallel to the (a) [100] and (b) [110] directions.
Open circles (plus) are strong (weak) resonance signals obtained from the measurements in static fields at 1.6~K,
while solid circles (cross) show strong (weak) resonance signals in pulsed fields at 1.4~K.
The solid lines represent the dipolar resonance modes from the multiboson spin-wave theory, using the following set of parameters: $g_c = 2.23$, $g_{ab} = 2.28$,
$J = 49.1$~GHz, $J_z = 45.0$~GHz, $\Lambda = 49.7$~GHz and $W_z^2J_{\it pz}=0.05$~GHz (see Sec.~\ref{sec:ESR_dynamics_fit} and Appendix~\ref{app:multib}).
The red dashed line in (a) indicates a resonance mode with a slope twice larger than the others, corresponding to a two-magnon excitation. 
 Yellow shaded area marks the phase below saturation for in-plane magnetic fields. 
}
\end{center}
\end{figure}
%-----------------------------------------------------------------------------------------------------------------------
%------      FIG 5    ---------------------------------------------------------------------------------------
\begin{figure}[ht!]
\begin{center}
\includegraphics[width=0.8\columnwidth]{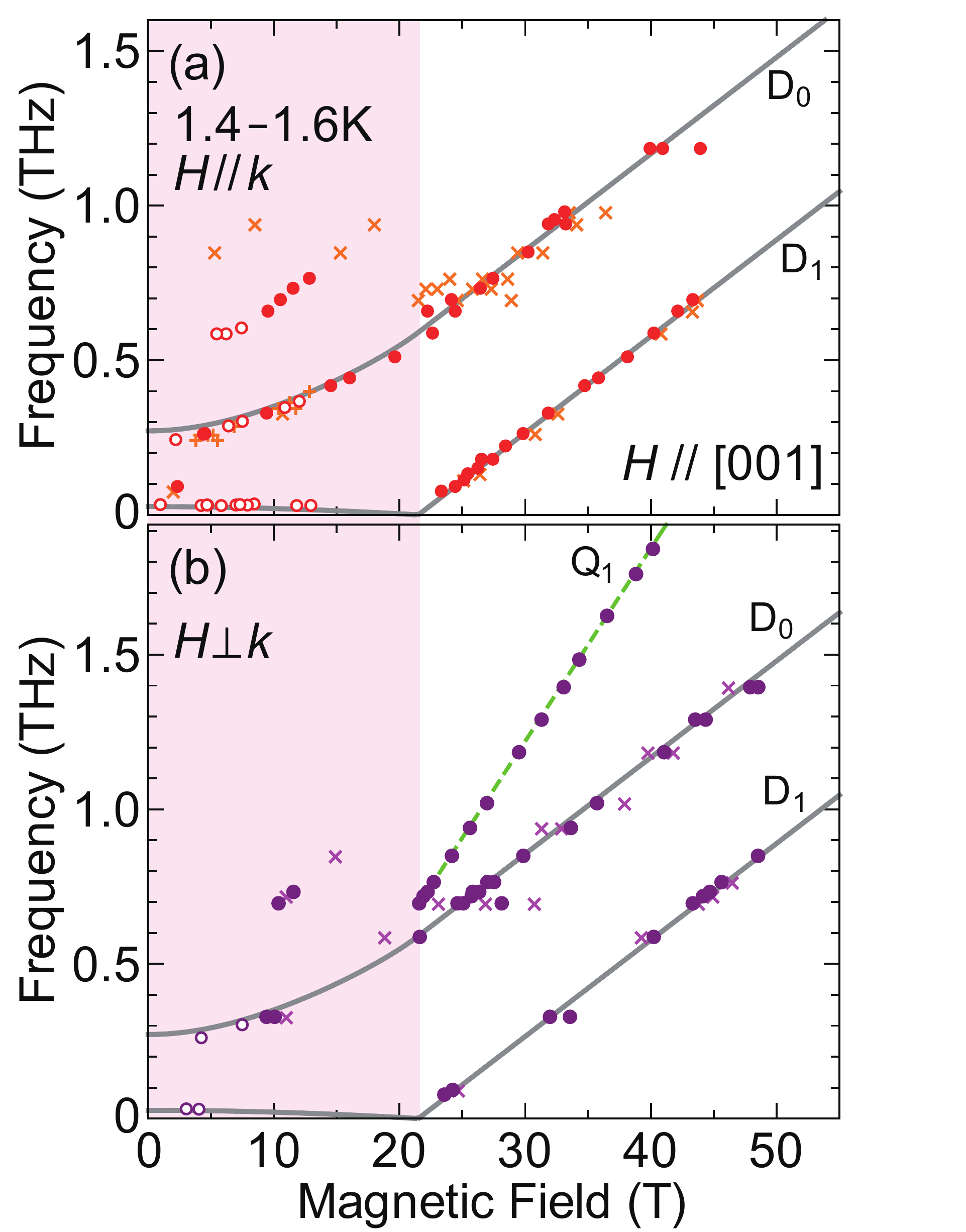}
\vspace{-9pt}
\caption{\label{fig:ESR001}  Frequency-field diagrams of the ESR resonance fields of Sr$_2$CoGe$_2$O$_7$ for magnetic fields parallel to the [001] directions in the (a) Faraday and (b) Voigt configuration, showing the two-magnon absorptions (green dashed line).
Open circles (plus) are strong (weak) resonance signals obtained from the measurements in static fields at 1.6~K,
while solid circles (cross) show strong (weak) resonance signals in pulsed fields at 1.4~K.
The solid lines represent the dipolar resonance modes calculated by the multiboson spin-wave theory with the same set of parameters as in Fig.~\ref{fig:ESR_100_110} (See Sec.~\ref{sec:ESR_dynamics_fit} and Appendix~\ref{app:multib}). The shaded area in red is the magnetic field range below the saturation field.
}
\end{center}
\end{figure}
%-----------------------------------------------------------------------------------------------------------------------

%%%%%%%%%%%%%%%%%%%%%%%%%%%%%%%%%%%%%%%%%%%%%%%%
\section{Dynamic properties studied by electron spin resonance}
\label{sec:ESR_dynamics}
%%%%%%%%%%%%%%%%%%%%%%%%%%%%%%%%%%%%%%%%%%%%%%%%

We continue our study with the dynamical properties. 
The ESR measurements have been performed for two configurations. 
In the Faraday configuration, the exciting electromagnetic wave propagates with wave vector $\mathbf{k}\propto \mathbf{E}^\omega\times\mathbf{H}^\omega$ in the direction of the external magnetic field ($\mathbf{H}\|\mathbf{k}$), consequently we can observe excitations driven by the components of oscillating electromagnetic field which are perpendicular to the external magnetic field.
In the Voigt configuration, the electromagnetic wave propagates in a perpendicular direction with respect to the external magnetic field ($\mathbf{H}\bot\mathbf{k}$), and the excitation spectrum contains electric and magnetic transitions coming from the fields of the light oscillating both parallel and perpendicular to the applied  field. 
Therefore, above the saturation, when the spins are aligned with the field, the transitions induced by the perpendicular component (with respect to the spin orientation) of the light are present in both Faraday and Voigt configuration, while the transitions induced by the parallel components are present in the Voigt configuration only. 
In both configurations we measured the ESR spectra for external magnetic fields parallel to $[100]$, $[110]$ and $[001]$ crystallographic directions using unpolarized light, having in total six different geometries.  
Measurements in different setups allow for the experimental verification of selection rules and identification of the observed excitations.

%%%%%%%%%%%%%%%%%%%%%%%%%%%%%
\subsection{Frequency--magnetic field plots of the resonance fields}
%%%%%%%%%%%%%%%%%%%%%%%%%%%%%

Figures~\ref{fig:ESR_100_110} and~\ref{fig:ESR001}(a) shows the frequency-magnetic field plots of ESR resonance fields at $1.4$ -- $1.6$ K in the Faraday configuration ($\mathbf{H}\|\mathbf{k}$). 
Near zero-field, three energy gaps are clearly identified: $\Delta_1 \approx$ 30~GHz, $\Delta_2 \approx$ 220~GHz, and $\Delta_3 \approx$ 700~GHz.
The gaps close to $\Delta_2$ and $\Delta_3$ were also reported in THz spectroscopy measurements of Ba$_2$CoGe$_2$O$_7$~\cite{Kezsmarki2011,PencPRL}, corresponding to transversal and longitudinal spin excitations, respectively. 
We can trace the origin of $\Delta_2$ to the interplay of exchange interaction and single-ion easy-plane anisotropy and $\Delta_3$ to the exchange interaction~\cite{Matsumoto2007, MiyaharaJPSJ, PencPRL}.

The magnetic field dependence of the excitations below 500~GHz clearly resembles the usual ESR spectra for the easy-plane antiferromagnets~\cite{Nagamiya1955}.
The only difference is the emergence of the smallest gap, $\Delta_1$.
This additional gap $\Delta_1$ is the result of a small anisotropy that fully breaks the spin rotational symmetry --- also breaking the remaining O(2) symmetry within the easy plane --- and can be explained by introducing polarization-polarization interaction which is also responsible for the vanishing of induced $P^z$ as the magnetic field approaches zero~\cite{RomPRB}.  
Similar energy gap has been observed in inelastic neutron diffraction measurements of Ba$_2$CoGe$_2$O$_7$~\cite{SodaPRL}.

%------      FIG 6    ---------------------------------------------------------------------------------------
\begin{figure}
\begin{center}
\includegraphics[width=0.9\columnwidth]{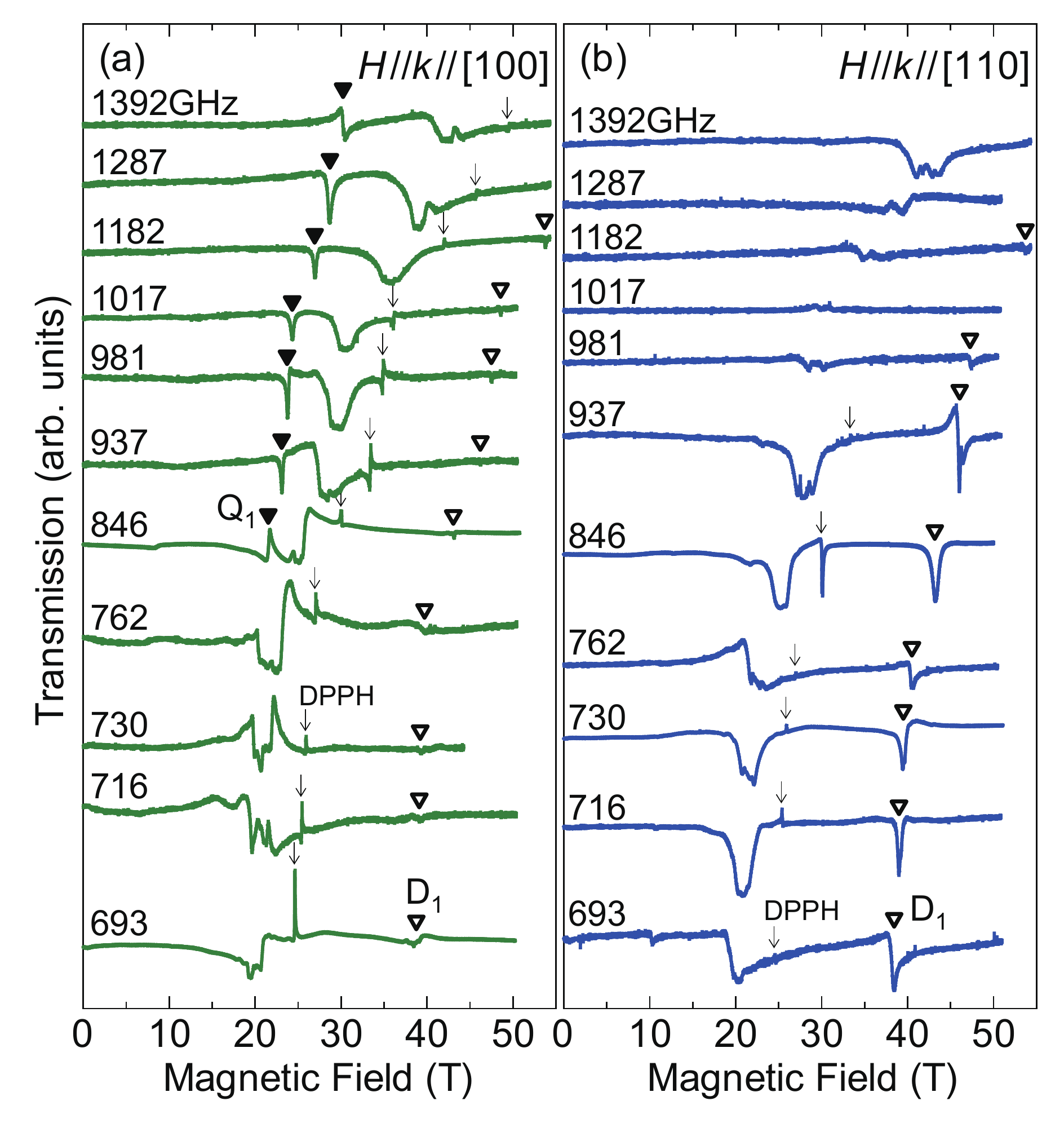}
\vspace{-9pt}
\caption{\label{fig:ESR_absorption_100_110} Frequency dependent ESR absorption spectra of SCGO in Faraday configuration at 1.4~K for (a) $H\| {\rm [100]}$ and (b) $H\| {\rm [110]}$.
Arrows mark the resonance fields of DPPH (2,2-diphenyl-1-picrylhydrazyl, ESR marker with $g=2.0036$).
Open and solid triangles mark the resonance fields of one-magnon (${\sf D}_1$) and two-magnon (${\sf Q}_1$) resonance modes, respectively. For Faraday configuration and the two in-plane static field direction, the ${\sf D}_1$ dipolar and 
${\sf Q}_1$ quadrupolar transitions appear to be mutually exclusive.}
\end{center}
\end{figure}
%-----------------------------------------------------------------------------------------------------------------------
%

Above the saturation, the modes are linearly increasing with the magnetic field, providing further evidence for a smaller anisotropy. We can identify the conventional magnon modes, ${\sf D}_0$ and ${\sf D}_1$ with the slopes corresponding to the $g$ factors we obtained from the magnetization measurements discussed in Sec.~\ref{sec:magnetization}. In the case of $\mathbf{H}\|[100]$, however, an additional mode, ${\sf Q}_1$, emerges at higher frequencies, its frequency increasing twice as fast with magnetic field as the frequency of the dipolar transitions, ${\sf D}_0$ and ${\sf D}_1$. This rapidly increasing mode corresponds to the absorption of two magnons.  

Such two-magnon excitation can be considered as a quadrupolar fluctuation. As discussed in Sec.~\ref{sec:Ham_ME}, the polarization is expressed by spin-quadrupole operators which may create two magnons on-site, a $\Delta S^z=2$ process. Here we argue that the ${\sf Q_1}$ two-magnon mode becomes visible in the ESR spectrum due to the quadrupole transitions driven by the oscillating electric field of the light, even when ${\sf Q_1}$ does not have dipole component but is purely quadrupolar.

Figure~\ref{fig:ESR_absorption_100_110} displays the ESR absorption spectra of SCGO at 1.4~K\@ in Faraday configuration.
The ${\sf Q}_1$ two-magnon mode is visible for $H \| {\rm [100]}$ only, as observed when comparing Fig.~\ref{fig:ESR_100_110} (a) and (b). 
The signal intensities of the ${\sf D}_1$ mode show a negligible absorption for $H \| {\rm [100]}$ and a strong absorption for $H \| {\rm [110]}$. 
It appears that the absorptions by the ${\sf D}_1$ and ${\sf Q}_1$ modes are mutually exclusive for these setups.

\subsection{Calculation of the one- and two-magnon spectra in high magnetic fields}  

The solid lines in Figures~\ref{fig:ESR_100_110} and~\ref{fig:ESR001} represent the resonance modes obtained by the multiboson spin-wave theory, which we discusse in details in the Appendix~\ref{app:multib}. This approach is suitable to treat the dipole and quadrupole type of excitations on an equal footing in the entire magnetic field regime, reproducing the correct  gaps. However, as we would like to focus on the quadrupole excitation, we introduce a simpler, more transparent model which works in the saturated phase, where the quadrupole mode was observed. 
To exhaust the possibilities of creating quadrupolar states, we calculate the two-magnon spectrum, which allows for the creation of two magnons at different sites. These magnons can then interact with each-other and even hop on the same site to form an onsite two-magnon excitation. This calculation nicely complements the multiboson spin wave approach, which can only capture the onsite two-magnon excitations, and is necessary to understand why the quadrupole mode coupling to the uniform polarization (${\sf Q}_0$) remains silent in the experiment.

In what follows, we present our analytical results for the excitation energies in high magnetic fields above the saturation. For simplicity we  neglect the $J_{pz}$ and $g_s$ terms in the Hamiltonian (\ref{eq:Hamiltonian}), since we expect them to be very small.
We note that supplemental material contains analytical solutions of a variational approach and multiboson spin-wave theory for the entire spectra, including the low field regime as well. 

%------      FIG 7    ---------------------------------------------------------------------------------------
\begin{figure}[b]
 \centering
 \includegraphics[width=0.95\columnwidth]{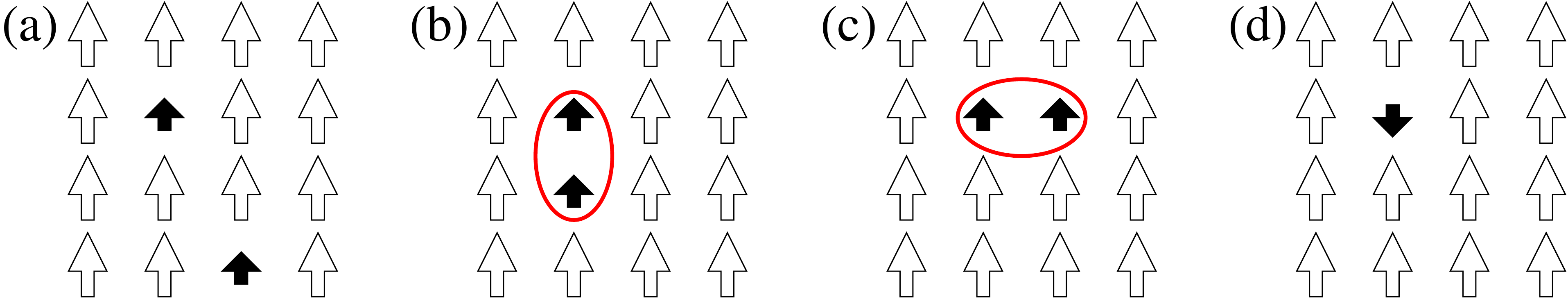}
\caption{(a) A typical two-magnon configuration, short black arrows show the $S^z = 1/2$ spins moving on the $S^z =3/2$ spin background (empty arrows). The $\mathbf{q}=(\pi,\pi)$ linear combinations  of the neighboring $S^z = 1/2$ spin states shown in (b)  and (c) are exact eigenstates, and they make the twofold degenerate ${\sf B}_1$ bound state.  (d) The down-pointing short arrow represents a $S^z=-1/2$ state; the  $\mathbf{q}=(\pi,\pi)$ linear combination of such states is the ${\sf Q}_1$  mode (see Eq.~(\ref{eq:Q1wavefunction})), which is also an eigenstate for the model with nearest neighbor exchanges only. 
\label{fig:2magnons}}
\end{figure}
%-----------------------------------------------------------------------------------------------------------------------

%=================================================================
\subsubsection{Excitations for the magnetic field $\|[001]\|z$}

\begin{figure}[b]
 \centering
 \includegraphics[width=0.8\columnwidth]{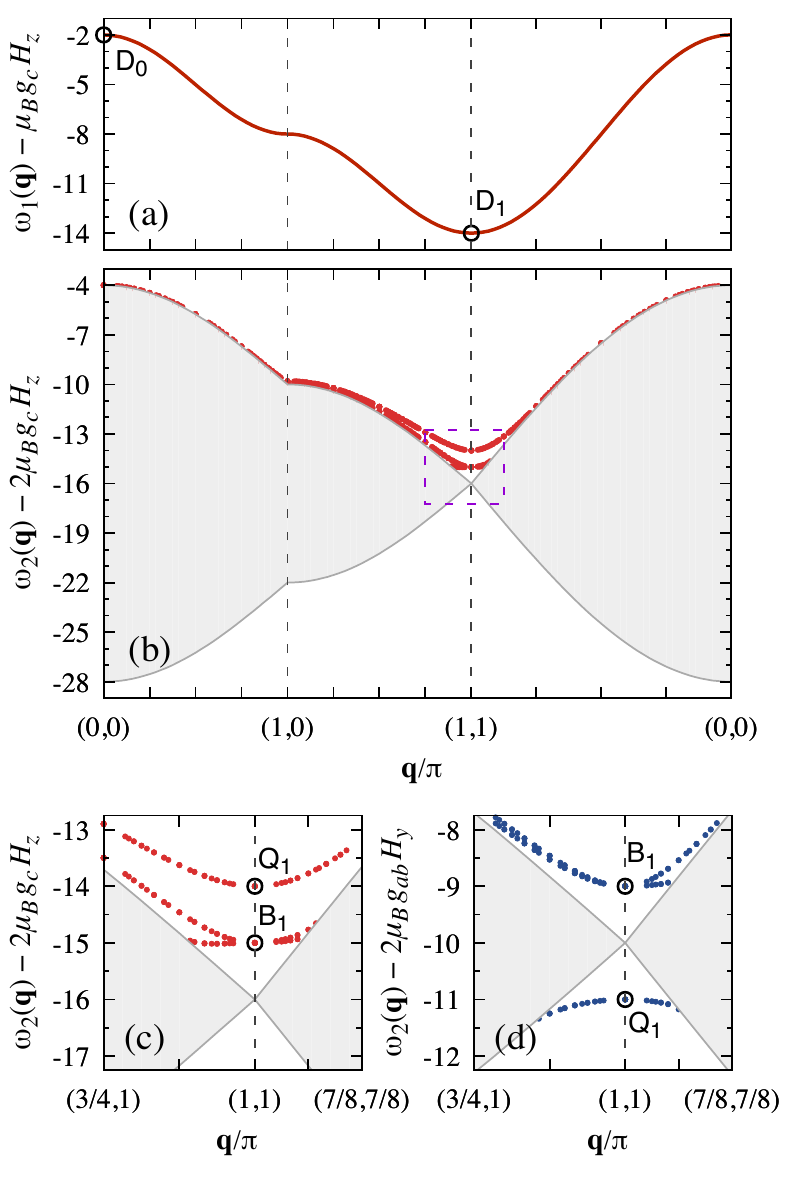}
\caption{One-- and two--magnon spectra in magnetic field above the saturation for $ \Lambda =1$ and $J=J_z=1$. (a) The dispersion of a single magnon along a path in the 2-dimensional Brillouin zone, following Eq.~(\ref{eq:1magnon_hz}). (b) In the two-magnon spectrum, the gray shaded area is the two-magnon continuum, the red points outside the continuum are the antibound states obtained from diagonalizing the two-magnon problem on different size clusters. (c) and (d) shows the magnified part of the two magnon spectrum close to $\mathbf{q}=(\pi,\pi)$.  The field is along the $[001]$ direction in the (a),(b), and (c), while it lies within the easy (001) plane in (d). The circles show the energy of the ${\sf D}_0$ and ${\sf D}_1$ mode in (a) and the energy of the ${\sf B}_1$
and ${\sf Q}_1$ mode in (c) and (d).
\label{fig:magnon_dispersion}}
\end{figure}

When the external magnetic field is parallel to the $[001]$ direction ({\it i.e.} perpendicular to the $(001)$ easy plane), the Hamiltonian shown in Eq.~(\ref{eq:Hamiltonian}) conserves the number of magnons. The ground state above the saturation is the trivial product of the $|S^z=3/2\rangle$ states over all the sites.
We need to invest $g_c \mu_{\rm B} H_z -2  \Lambda - 6  J_z $ diagonal energy to create a single magnon, the $|S^z=1/2\rangle$ spin state on a site $j$ with the $\hat{S}^-_j$ operator. This spin state can then hop with amplitude $3J/2$ to neighboring sites.
The resulting state is described by the 
\begin{align}
 |{\sf{D}}(\mathbf{q}) \rangle &=
 \sum_j e^{i \mathbf{q}\cdot\mathbf{r}_j } \hat{S}^-_j | g_s  \rangle
\end{align}
wave function, with energy  
\begin{align}
\omega_{1}(\mathbf{q}) &=
 \mu_{\rm B} g_c H_z - 2  \Lambda - 6 J_z + 6 J \gamma_{\mathbf{q}}  \; ,
  \label{eq:1magnon_hz}
\end{align}
where $\gamma_{\mathbf{q}}$ is the geometrical factor $\frac{1}{2}(\cos q_x + \cos q_y)$. We plot the dispersion above in Fig.~\ref{fig:magnon_dispersion}(a) for a realistic ratio of parameters $J=J_z = \Lambda$, which is very close to the one we will get for SCGO below, in Sec.~\ref{sec:ESR_dynamics_fit}.

Due to the alternation of the tetrahedra, the unit cell is doubled, so magnons at both  $\mathbf{q} = (0,0)$ and $\mathbf{q} = (\pi,\pi)$ wave vectors in the extended Brillouin zone are excited in the ESR spectrum (note that the unit cell of the Hamiltonian contains only one spin), with energy
\begin{eqnarray}
\omega_{\sf{D}_0}&\equiv& \omega_{1}(0,0) = \mu_{\rm B} g_c H_z  -2 \Lambda -6 J_z + 6 J \;,\\
\omega_{{\sf D}_1}&\equiv& \omega_{1}(\pi,\pi) =  \mu_{\rm B} g_c H_z  -2 \Lambda -6 J_z -6 J \;.\label{eq:D1_z}
\end{eqnarray}
The energy difference between these two modes is simply the bandwidth of the magnons, equal to $12J$.

%------      FIG 8    ---------------------------------------------------------------------------------------
%-----------------------------------------------------------------------------------------------------------------------

Two magnons can propagate freely with the dispersion given above, except when they meet at neighboring sites as shown in Figs.~\ref{fig:2magnons}(b) and (c). Furthermore, the two magnons can hop onto each other, creating an $|S^z=-1/2\rangle$ spin state on a single site, Fig.~\ref{fig:2magnons}(d). This leads to an interaction between the magnons, and we need to solve the corresponding two-body problem. While this can be done analytically 
\cite{Wortis,Hanus,Mattisbook,Silberglitt1970,Zhitomirsky2010}, here we resort to exact diagonalization in the two-magnon Hilbert space, and plot the spectrum for finite clusters. In Fig.~\ref{fig:magnon_dispersion}(b), we show the result compiled from different size clusters (up to 3200 sites), which respect the full $D_4$ symmetry of the square lattice. The propagating magnons form a two magnon continuum. The continuum is the broadest at the center of the Brillouin zone, where its width of $24 J$ is twice the bandwidth of the magnons, and  shrinks to a single, highly degenerate point at the $\mathbf{q} = (\pi,\pi)$ corner of the Brillouin zone, with energy 
\begin{align}
  \omega_2^{\text{cont.}}(\pi,\pi) = 2 \mu_{\rm B} g_c H_z  - 4  \Lambda - 12 J_z \;. 
\end{align}
We observe that close to $\mathbf{q} = (\pi,\pi)$, three distinct state split off from the two-magnon continuum due to the interaction. In fact, at $\mathbf{q} = (\pi,\pi)$ the 
\begin{align}
  |{\sf Q}_1\rangle = \sum_j (-1)^j \hat{S}^-_j \hat{S}^-_j | g_s \rangle
  \label{eq:Q1wavefunction} 
\end{align}
state (shown in Fig.~\ref{fig:2magnons}(d)), with energy 
\begin{align}
 \omega_{\sf{Q}_1} 
 %\equiv \omega_2^{|-1/2\rangle}(\pi,\pi) 
 = 2 \mu_{\rm B} g_c H_z - 2  \Lambda - 12 J_z  \;, 
 \label{eq:Q1_z}
\end{align}
 is decoupled from the other states \cite{Silberglitt1970}. So are decoupled the configurations having two neighboring $|+1/2\rangle$ states in either of the directions [Fig.~\ref{fig:2magnons}(b) and (c)], with energy 
\begin{align}
  \omega_{\sf{B}_1} = 2 \mu_{\rm B} g_c H_z - 4  \Lambda - 11 J_z  \;. 
  \label{eq:2magnon_hz_11}
\end{align}
 The twofold degeneracy of this state is lifted as we go away from the $\mathbf{q} = (\pi,\pi)$ point \cite{Wortis,Hanus}, as shown in Fig.~\ref{fig:magnon_dispersion}(c).

\subsubsection{Excitations for the external magnetic field in the (001) easy plane (plane perpendicular to the $z$ axis)}

When the field is along the $[100]$ or $[110]$ direction, the Zeeman term and the exchange term in Hamiltonian~(\ref{eq:Hamiltonian}) do not commute any more. Instead, we can resort to perturbation expansion.
Assuming a large magnetic field in the $y$ direction, we introduce a new basis for the spin operators $ (\hat{S}^x, \hat{S}^y, \hat{S}^z) \rightarrow (\tilde S^y, \tilde S^z, \tilde S^x)$, so that the quantization axis is along the magnetic field. We can then rewrite the Hamiltonian as $\mathcal{H} = \mathcal{H}_0 + \mathcal{H}_1$, where $\mathcal{H}_0$ is diagonal in the new basis, 
\begin{align}
\mathcal{H}_0 &= 
\frac{ \Lambda  }{2} \sum_i \left(\frac{15}{4}- {\tilde S}^{z}_{i} {\tilde S}^{z}_{i} \right)
+ J \sum_{(i,j)}{\tilde S}^{z}_{i} {\tilde S}^{z}_{j}
\nonumber\\
& - \mu_{\rm B} H_y g_{ab} \sum_i {\tilde S}^{z}_{i} \;,
\end{align}
and $\mathcal{H}_1$ contains the off-diagonal matrix elements,
\begin{align}
\mathcal{H}_{1} &= 
  \frac{ J_z + J}{4} \sum_{(i,j)} \left( {\tilde S}^{-}_{i} {\tilde S}^{+}_{j} + {\tilde S}^{+}_{i} {\tilde S}^{-}_{j} \right) 
 \nonumber\\&
+\frac{ \Lambda }{4} \sum_{i} \left({\tilde S}^{-}_{i}{\tilde S}^{-}_{i} + {\tilde S}^{+}_{i}{\tilde S}^{+}_{i} \right) 
\nonumber\\
&
+\frac{ J_z - J}{4} \sum_{(i,j)} \left( {\tilde S}^{-}_{i} {\tilde S}^{-}_{j} + {\tilde S}^{+}_{i} {\tilde S}^{+}_{j} \right) \;.
\end{align}
(the spectra do not depend on the actual direction of the field in the $xy$ plane, as the anisotropy has an $O(2)$ symmetry about the $z$ axis when $J_{pz}=0$)

Below we perform a first order degenerate perturbation expansion using the $\mathcal{H}_1$ as a perturbation operator. The ground state of the  $\mathcal{H}_0$ operator is a ferromagnetic state with all the spins pointing along the field.  
%Since $ \Lambda $ appears both in $\mathcal{H}_0$ and $\mathcal{H}_1$, to get simpler expressions we will expand the result in $ \Lambda $. 
The  ${\tilde S}^{-}_{i} {\tilde S}^{-}_{i}$ and ${\tilde S}^{-}_{i} {\tilde S}^{-}_{j}$ processes, which change the number of magnons, first appear in the second order of perturbation expansion, and we neglect them in the following, assuming a small single-ion anisotropy and nearly isotropic exchange. Within this approximation, the problem is equivalent with the one we discussed in the previous subsection for $\mathbf{H}\|[001]$, but we need to replace $J$ by $(J+J_z)/2$, $J_z$ by $J$, and the on site anisotropy $\Lambda$ by $-\Lambda/2$ in all of Eqs.~(\ref{eq:1magnon_hz})-(\ref{eq:2magnon_hz_11}). Most notably, the sign of the effective anisotropy changes. The dispersion of single magnon then becomes
\begin{align}
\omega_{1}(\mathbf{q}) &=
  \mu_{\rm B} g_{ab} H_y +  \Lambda - 6 J  +\left(3 J+ 3 J_z \right) \gamma_{\mathbf{q}} + \cdots \;,
 \label{eq:1magnon_hx}
\end{align}
where the dots denote the neglected second and higher order terms.
Continuing the replacements, we get
\begin{subequations}
\begin{align}
\omega_{{\sf D}_0} &=  \mu_{\rm B} g_{ab} H_y  + \Lambda - 3 J + 3 J_z + \cdots \;,
\label{D0_xy}
\\
\omega_{{\sf D}_1} &=  \mu_{\rm B} g_{ab} H_y + \Lambda - 9 J -  3 J_z +  \cdots \;,\label{eq:D1_xy}
\\
\omega_2^{\text{cont.}}(\pi,\pi) &= 2 \mu_{\rm B} g_{ab}  H_y + 2  \Lambda - 12 J + \cdots  \;,
\\
\omega_{\sf{Q}_1} & =  2 \mu_{\rm B} g_{ab}  H_y +  \Lambda - 12 J + \cdots  \;, \label{eq:Q1_x}
\\
\omega_{\sf{B}_1} &= 2 \mu_{\rm B} g_{ab}  H_y + 2  \Lambda - 11 J + \cdots \;.
\end{align}
\end{subequations}
The ${\sf{Q}_1}$ is now below the continuum (for $\Lambda>0$), as shown in Fig.~\ref{fig:magnon_dispersion}(d), forming a bound state.

\begin{figure*}
\begin{center}
\includegraphics[width=1.9\columnwidth]{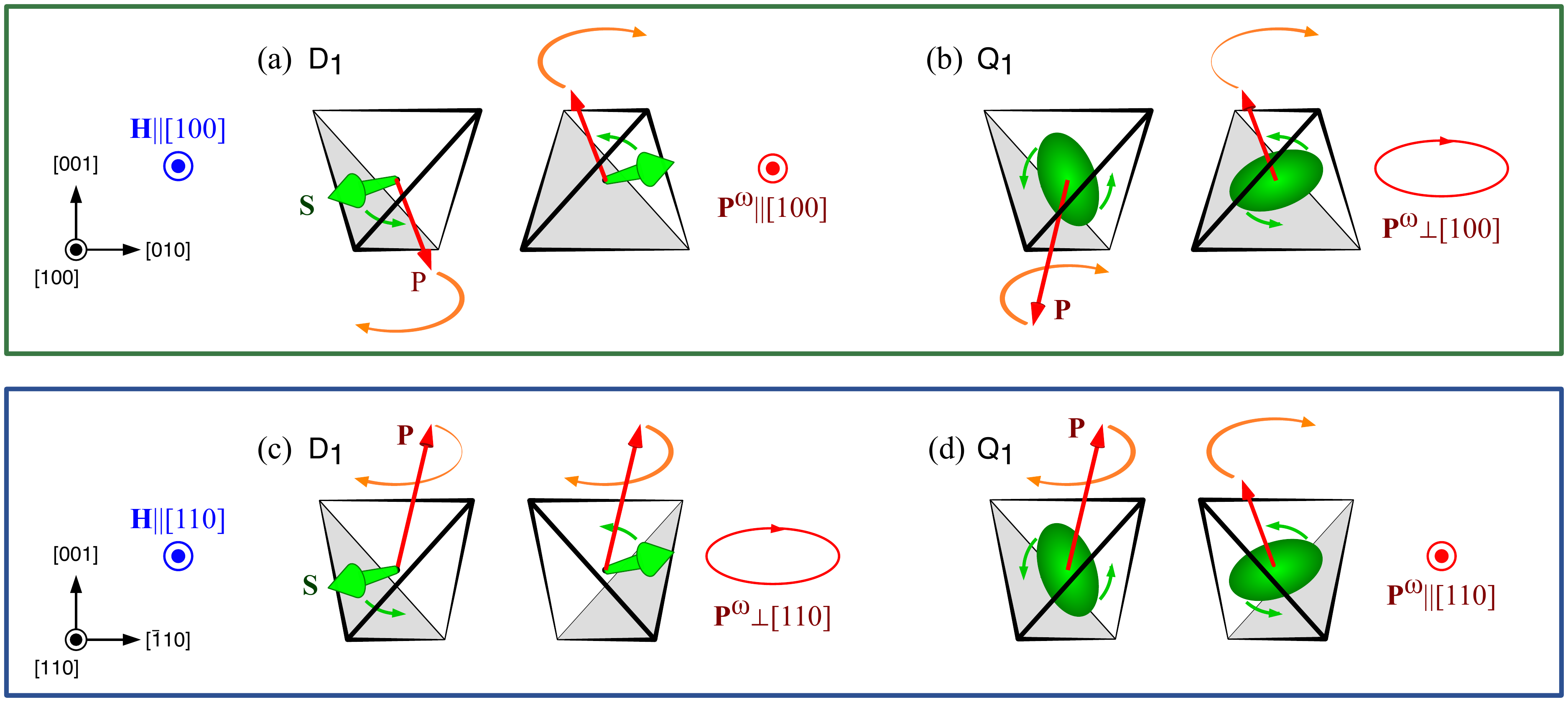}
\vspace{-9pt}
\caption{\label{fig:schematic_waves} 
  Schematic plot of the dipolar [(a) and (c)] and quadrupolar [(b) and (d)] modes in different geometries, as seen from the direction of the magnetic field. 
In the dipolar wave ${\sf D}_1$  for $\mathbf{H}\|[110]$ (c) and the quadrupolar mode ${\sf Q}_1$ for $\mathbf{H}\|[100]$ (b) the oscillating component of the uniform electric polarization $\mathbf{P}^\omega$ (shown by red ellipse) is perpendicular to the external magnetic field $\mathbf{H}$, therefore they are active in the Faraday configuration. 
In the ${\sf D}_1$ for $\mathbf{H}\|[100]$ (a) and ${\sf Q}_1$ for $\mathbf{H}\|[110]$ (d) the $\mathbf{P}^\omega \| \mathbf{H}$ ({\it i.e.} it oscillates in and out from the shown plane), so these modes are active in Voigt configuration only.  The green ellipse represents the rotating quadrupolar moments, while the green arrows the precessing dipolar spins on the two sublattices. The red arrows show the electric polarization vectors which are excited by the oscillating electric field. Animations of these modes are shown in Supplement~\cite{movies}
}
\end{center}
\end{figure*}

The difference between the dipolar and quadrupolar modes is illustrated schematically in Fig.~\ref{fig:schematic_waves}. Denoting  the spin component parallel to the external field by $S^{||}$ and the two orthogonal components  by $S^{\perp_1}$ and $S^{\perp_2}$,  the expectation value of the spin operators are
\begin{align}
 \langle \hat{S}^{\perp_1} \rangle &\propto (-1)^j \sin(\omega_{{\sf D}_1} t) \;,
 \nonumber\\
 \langle \hat{S}^{\perp_2}_j \rangle &\propto (-1)^j \cos(\omega_{{\sf D}_1} t) \;,
 \nonumber\\
 \langle \hat{S}^{||}_j \rangle &\approx \frac{3}{2}\; 
\end{align} 
in the ${\sf D}_1$ dipolar mode.
They describe the usual precession of dipolar components of the spin [green arrows in Fig.~\ref{fig:schematic_waves}(a) and (c)] around the magnetization axes. In the ${\sf Q}_1$ quadrupolar mode transversal components of the spin are $\langle S^{\perp_1} \rangle = \langle S^{\perp_2} \rangle = 0$, instead the   
% \begin{align}
% %$\langle S^{[001]}_j S^{[010]}_j + S^{[010]}_j S^{[001]}_j \rangle \propto \sin(\omega_{{\sf Q}_1} t + \phi_j)$ 
%\langle \overline{S^{\perp_1}_j S^{\perp_2}_j} \rangle &\propto (-1)^j \sin(\omega_{{\sf Q}_1} t) \;,
%    \nonumber\\
%   \langle (S^{\perp_1}_j)^2-(S^{\perp_2}_j)^2 \rangle &\propto (-1)^j \cos(\omega_{{\sf Q}_1} t) \;
%\end{align}
\begin{align}
\langle \hat{Q}^{2\perp_1\perp_2}_j \rangle &\propto (-1)^j \sin(\omega_{{\sf Q}_1} t) \;,
    \nonumber\\
   \langle \hat{Q}^{\perp_1^2-\perp_2^2}_j \rangle &\propto (-1)^j \cos(\omega_{{\sf Q}_1} t) \;
\end{align}
components of the quadrupolar moment [shown as a green ellipse in Fig.~\ref{fig:schematic_waves}(b) and (d)]
rotate around the static dipolar moment aligned with the external magnetic field.

%------      FIG -6    ---------------------------------------------------------------------------------------
%\begin{figure}
%\begin{center}
%\includegraphics[scale=0.48, clip]{schematic_wave_V5.pdf}
%\vspace{-9pt}
%\caption{\label{fig6} 
%  Schematic plot of the (a) quadrupolar mode ${\sf Q}_1$ for $\mathbf{H}\|[100]$ and (b) the dipolar wave ${\sf D}_1$  for $\mathbf{H}\|[110]$, as seen from the direction of the magnetic field. (a) The green ellipse represents the quadrupolar moments $\langle (S^{[001]}_j)^2-(S^{[010]}_j)^2 \rangle \propto \cos(\omega_{{\sf Q}_1} t + \phi_j)$ and 
%%$\langle S^{[001]}_j S^{[010]}_j + S^{[010]}_j S^{[001]}_j \rangle \propto \sin(\omega_{{\sf Q}_1} t + \phi_j)$ 
%$\langle \overline{S^{[001]}_j S^{[010]}_j} \rangle \propto \sin(\omega_{{\sf Q}_1} t + \phi_j)$rotating around the $(\langle S^{[100]} \rangle,\langle S^{[010]} \rangle,\langle S^{[001]} \rangle)  \approx (3/2,0,0)$ static spin.   (b) The ${\sf D}_1$ mode is a usual spin--wave mode, where the $\langle S^{[\bar110]} \rangle \propto \sin(\omega_{{\sf D}_1} t + \phi_j)$ and $\langle S^{[001]}_j \rangle \propto \cos(\omega_{{\sf D}_1} t + \phi_j)$ (green arrows). The red arrows show the electric polarization vector. The $\phi_j = j\pi$ phase ensures the $\mathbf{q}=(\pi,\pi)$ wave vector of the excitations.
%  }
%\end{center}
%\end{figure}

%------ FIG 9 --------------------------------------------------------------------

%-----------------------------------------------------------------------------------------------------------------------

%%%%%%%%%%%%%%%%%%%%%%%%%%%%%%%%%%%%%%%%%%%%%%%%
\subsection{Fitting the parameters}
\label{sec:ESR_dynamics_fit}
%%%%%%%%%%%%%%%%%%%%%%%%%%%%%%%%%%%%%%%%%%%%%%%%

\begin{figure}[bt]
\begin{center}
\includegraphics[width=0.7\columnwidth]{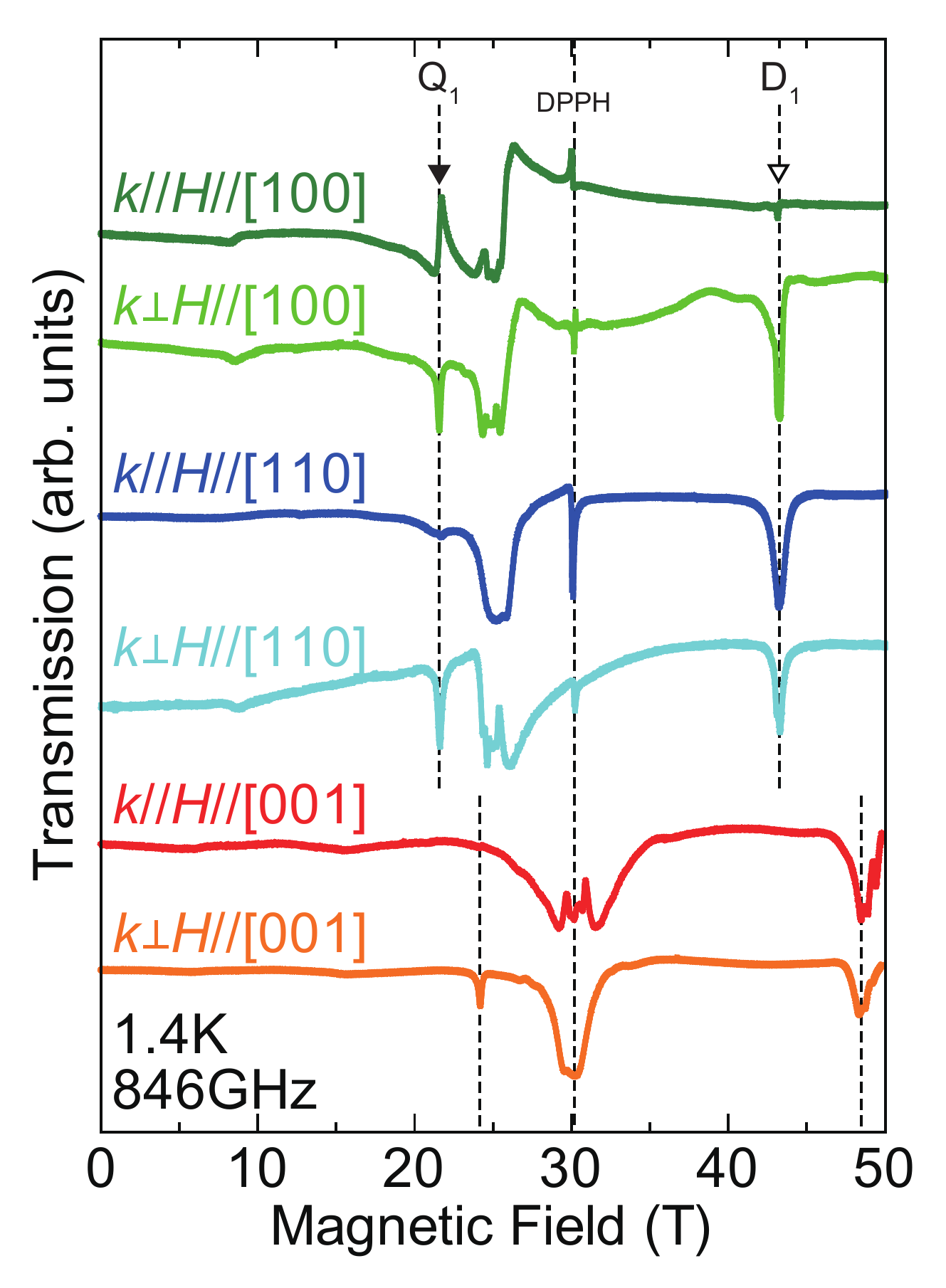}
\caption{\label{fig:ESR846GHz} 
ESR absorption spectra of Sr$_2$CoGe$_2$O$_7$ at 846~GHz in Faraday ($\mathbf{k}\|\mathbf{H}$) and Voigt ($\mathbf{k}\perp\mathbf{H}$) configuration for $\mathbf{H}\|[100]$, $\mathbf{H}\|[110]$, and $\mathbf{H}\|[001]$.
Vertical dotted lines, from left to right, indicate two-magnon resonance signal, signal of ESR standard DPPH, and one-magnon resonance signal, respectively. 
}
\end{center}
\end{figure}

We will use the modes linear in magnetic field above saturation in Figs.~\ref{fig:ESR_100_110} and \ref{fig:ESR001} to extract parameters of the Hamiltonian. 
The $\omega_{{\sf D}_1}$ and $\omega_{{\sf Q}_1}$ modes show a narrow absorptions in Figs.~\ref{fig:ESR_absorption_100_110} and \ref{fig:ESR846GHz}, while the $\omega_{{\sf D}_0}$ resonance mode is broad, giving us a natural choice of selecting the narrow $\omega_{{\sf D}_1}$ (solid circles in Fig.~\ref{fig:ESR_100_110}(b) for $\mathbf{H}\|[110]$ and \ref{fig:ESR001}(a) for $\mathbf{H}\|[001]$) and $\omega_{{\sf Q}_1}$ modes (solid circles in Fig.~\ref{fig:ESR_100_110}(a) for $\mathbf{H}\|[100]$ and \ref{fig:ESR001}(b) for  $\mathbf{H}\|[001]$) to determine the precise values of the $ \Lambda$, $J$, $J_z$, and the $g$ values. 

Performing a standard multiple linear regression fit using Eqs.~(\ref{eq:D1_z}), (\ref{eq:Q1_z}), (\ref{eq:D1_xy}), and (\ref{eq:Q1_x}) as a model, we get the following parameters for the $g$ values,
%g_{\sf{D}_1}^{[001]} &= 2.23 \pm 0.02 \;,\\
%g_{\sf{D}_1}^{[110]} &= 2.28 \pm 0.01 \;,\\
%g_{\sf{Q}_1}^{[001]} &= 4.42 \pm 0.02 \;,\\
%g_{\sf{Q}_1}^{[100]} &= 4.57 \pm 0.02 \;,\\
%J &= 49.1 \pm 0.8~\text{GHz} = 2.36 \pm 0.04~\text{K} \;,\\
%J_z &= 45.0 \pm 1.1~\text{GHz} = 2.21 \pm 0.05~\text{K} \;,\\
%\Lambda &= 49.7 \pm 4.2~\text{GHz} = 2.39 \pm 0.20~\text{K} \;.
%\end{align}
\begin{subequations}
\begin{align}
g_{\sf{D}_1}^{[001]} &= 2.23 \pm 0.02 \;, \label{eq:gD1[001]fit} \\
g_{\sf{Q}_1}^{[001]} &= 4.42 \pm 0.02 \;,\\
g_{\sf{D}_1}^{[110]} &= 2.28 \pm 0.01 \;, \label{eq:gD1[110]fit} \\
g_{\sf{Q}_1}^{[100]} &= 4.57 \pm 0.02 \;,
\end{align}\label{eq:gs}
\end{subequations}
where, instead of $g_{ab}$ and $g_{c}$, we left the $g$-values of the corresponding modes as free parameters which measure the slope. The values of the exchange couplings and on--site anisotropy are
%\begin{align}
\begin{subequations}
\begin{align}
J &= 49.1 \pm 0.8~\text{GHz} = 2.36 \pm 0.04~\text{K} \;, \label{eq:Jfit} \\
J_z &= 45.0 \pm 1.1~\text{GHz} = 2.16 \pm 0.05~\text{K} \;, \label{eq:Jzfit} \\
\Lambda &= 49.7 \pm 4.2~\text{GHz} = 2.39 \pm 0.20~\text{K} \;. \label{eq:Lambdafit}
\end{align}\label{eq:JandLambda}
\end{subequations}

The exchange parameter $J$ is in perfect agreement with the measured shift between the two single-magnon modes, $\omega_{{\sf D}_0}-\omega_{{\sf D}_1} = 12J \approx 600$~GHz in Fig.~\ref{fig:ESR001}~\cite{conversion}. The fit gives an $\omega_{{\sf D}_1}=0$ intercept at $16.6\pm0.1$~T, while for $H\|[001]$ the intercept $21.5\pm0.3$~T coincides with the saturation field, $21.5$~T (Figs.~\ref{fig:staticpolarization}(a) and \ref{fig:ESR001}).

Along the $[100]$ the signal of the ${\sf D}_1$ mode is quite weak, and the points are more scattered around the line. Consistently, the error of a linear fit is larger, we get $g_{\sf{D}_1}^{[100]} = 2.27 \pm 0.04 $ for the slope and $-524 \pm  23~\text{GHz}$ for the zero field intercept. According to our theory, the dynamics of this weak mode is governed by Eq.~(\ref{eq:D1_xy}), like for the $\omega_{\sf{D}_1}^{[110]} $mode. Replacing the fitted values given by Eqs.~(\ref{eq:Jfit})-(\ref{eq:Lambdafit}), for the intercept we get $527~\text{GHz}$, which is well within the estimated error bars. Similarly, we can conclude, that $ g_{\sf{D}_1}^{[110]} = g_{\sf{D}_1}^{[001]}$ within error bars, {\it i.e.} the slope of the two modes is equal, and can be identified with the $g_{ab}$. We choose the more precise $g_{ab} =  g_{\sf{D}_1}^{[110]} =  2.28 \pm 0.01$ value.

The fit supports the two--magnon origin of the ${\sf Q}_1$ excitations, as $g_{\sf{Q}_1}^{[001]} \approx 2 g_{\sf{D}_1}^{[001]}$, so we can associate  $g_{\sf{D}_1}^{[001]}$  with $g_c$ and $g_{\sf{Q}_1}^{[001]}$ with $2 g_c$. Furthermore the $g_{{\sf Q}_1}^{[100]}$ is twice the $g_{{\sf D}_1}^{[100]}$ within the error bars.

%If we include all five lines
%\begin{align}
%g_{\sf{D}_1}^{[001]} &= 2.23 \pm 0.03 \;,\quad 
%g_{\sf{Q}_1}^{[001]} = 4.42 \pm 0.04 \;,\\
%g_{\sf{D}_1}^{[100]} &= 2.27 \pm 0.02 \;,\quad
%g_{\sf{Q}_1}^{[100]} = 4.57 \pm 0.04 \;,\\
%g_{\sf{D}_1}^{[110]} &= 2.27 \pm 0.02 \;,\\
%J &= 49.0 \pm 1.5~\text{GHz} = 2.35 \pm 0.07~\text{K} \;,\\
%Jz &= 45.8 \pm 2.2~\text{GHz} = 2.20 \pm 0.11~\text{K} \;,\\
%\Lambda &= 50.9 \pm 8.0~\text{GHz} = 2.44 \pm 0.38~\text{K} 
%\end{align}
%
 
 Finally, we numerically calculated the the single-magnon modes ${\sf D}_0$ and ${\sf D}_1$ from the multiboson spin-wave theory~\cite{juditPRB} using the above set of parameters. Adding a small $W_z^2J_{\it pz}/k_{\rm B}=2.4 \times 10^{-3}$~K, the calculated and the measured modes show an excellent agreement, even in the low field regime, as shown in Figs.~\ref{fig:ESR_100_110} and \ref{fig:ESR001}.

Neglecting the small $J_{\it pz}$, we performed the multiboson calculation analytically. We determined the energies of the ${\sf D}_0$ and ${\sf D}_1$ modes explicitly in Eqs.~(\ref{eq:D1_mode}) and~(\ref{eq:D0_modeZ}) of the Appendix~\ref{app:multib}. For zero magnetic field, we obtain
\begin{eqnarray}
%\Delta_2=(6J+\Lambda)\sqrt{2\left(1+\frac{J_z}{J}-\frac{12 J_z}{6J+\Lambda}\right)}\;,\
\Delta_2= \sqrt{24 J (3J - 3 J_z + \Lambda)}
\label{eq:Lambda_2}
\end{eqnarray}
in the leading order in anisotropies. 
Inserting the fitted parameter values of Eqs.~(\ref{eq:JandLambda}) we get $\Delta_2=270 \pm 30$~GHz. This is close to the experimentally observed gap, $\Delta_2 \sim 220$~GHz. The gap value for small anisotropies is very sensitive to the $\Lambda/J$ and $(J - J_z)/J$. Given that we have taken the $\Lambda$, $J$ and $J_z$ from the high field measurements, and the simplicity of the model, the correspondence is reasonable.

Quite interestingly, the anisotropy gap from the multiboson spin-wave theory, $\Delta_2 = \sqrt{24 J \Lambda}$ in the leading order and assuming $J=J_z$, is different from the standard linear spin wave calculation providing $\Delta_2 = \sqrt{16 J \Lambda S^2} = \sqrt{36  J \Lambda}$. In fact, going beyond the linear spin wave theory  quantum correction appear and the anisotropy gap becomes $\Delta_2 = \sqrt{16 J \Lambda S (S-1/2)} = \sqrt{24  J \Lambda}$ for $S=3/2$ \cite{Lindgard1976}, coinciding with the gap obtained from the multiboson spin-wave theory.

\subsection{Comparison to Ba$_2$CoGe$_2$O$_7$}
The Ba$_2$CoGe$_2$O$_7$ is a member of the \aa kermanite family, where the on-site anisotropy is believed to be large, estimates range from  $ \Lambda /J = 5.8$ \cite{PencPRL} to $ \Lambda /J \approx 8$ \cite{MiyaharaJPSJ}. The associated spin shortening is $\langle S \rangle \approx 1.3$ to $1.35$, about 10\% of the full spin value of 3/2 --- the magnetization at the saturation field is about 10\% smaller from the fully saturated value in very high magnetic field \cite{HutanuPRB}. In comparison, the spin shortening in SCGO is $\langle S \rangle = 1.491$, less than 1\%. This is why the magnetization curve shown in Fig.~\ref{fig:staticpolarization}(a) is so flat above 18~T\@.

The far-infrared absorption spectra of the Ba$_2$CoGe$_2$O$_7$ were studied in Ref.~\cite{PencPRL}. In comparison, the modes above the saturation field have a finite curvature -- the signature of the stronger anisotropy. The multiboson spin-wave described the Ba$_2$CoGe$_2$O$_7$ excitation spectrum with $\Lambda = 13.4$~K, $J=2.3$~K, $J_z = 1.8$~K --- the values of the exchange coupling are very similar, it is the single-ion anisotropy that is different in the two materials.

%------      FIG 9    ---------------------------------------------------------------------------------------

%%%%%%%%%%%%%%%%%%%%%%%%%%%%%
\section{Selection rules }
\label{sec:selection_rules}
%%%%%%%%%%%%%%%%%%%%%%%%%%%%%

To further characterize the high field excitations, we compared the ESR spectra for the six geometries in Fig.~\ref{fig:ESR846GHz}.
It appears that all the modes are active in the Voigt configurations. 
The only information about the matrix elements is provided by absence of the light absorptions in Faraday configurations. 
For example, in Fig.~\ref{fig:ESR846GHz} the ${\sf D}_1$ is not present for the $\mathbf{H}||[100]$ case, and the ${\sf Q}_1$ signal is missing when the external magnetic field is along the $[110]$ and $[001]$ directions. 
Since in the Faraday configuration the oscillating electric and magnetic fields of the incoming light are perpendicular to the direction of the external field, the missing absorption indicates that it is not excited by these components. In other words, it has no matrix elements with the perpendicular components of spin and polarization operators.  
Therefore, a signal present in Voigt, but absent in Faraday configuration, must be excited via the oscillating magnetic and/or electric fields which are parallel to the external field, and thus parallel to the magnetic moments above saturation.

\subsection{Selection rules observed in the experiment}

The situation discussed above happens in the  case $\mathbf{H} \| {\rm [100]}$, as the ${\sf D}_1$ magnon (open triangle in Fig.~\ref{fig:ESR846GHz}) is missing in the Faraday (dark green), but is present for the Voigt (light green) configuration. 
This means that only the [100] components of the light, $H^\omega_{[100]}$ or  $E^\omega_{[100]}$, excite ${\sf D}_1$. The ${\sf Q}_1$ two-magnon excitation (solid triangle in Fig.~\ref{fig:ESR846GHz}) is, however, present both in the Faraday and Voigt configuration, therefore it is coupled to the perpendicular [010] and/or [001] components. 

Similarly, when $\mathbf{H} \| {\rm [110]}$, the $[\bar110]$ and/or [001] components of the oscillating fields excite the ${\sf D}_1$ magnon present in Faraday and Voigt configurations, and the $[110]$ parallel components create the ${\sf Q}_1$ two-magnon missing in the Faraday configuration, but observed in the Voigt spectrum. Since we used unpolarized light, we cannot tell  whether the $E^\omega_{[110]}$ or/and $H^\omega_{[110]}$ excite the ${\sf D}_1$ magnon mode from the experiment.
 
Finally, when $\mathbf{H} \| {\rm [001]}$, it is the ${\sf Q}_1$ mode which appears in the Voigt configuration only, therefore it must couple to the parallel components $H^\omega_{[001]}$ and/or $E^\omega_{[001]}$, and only to those components of the incoming light. 

\subsection{Selection rules for the magnetic transitions}

Here we calculate the selection rules coming from the oscillating magnetic field $\mathbf{H}^\omega$ of the light, based on the  Zeeman coupling in the Hamiltonian, Eq.~(\ref{eq:Hamiltonian}).
Above saturation, all the spins are parallel to the external field, making a translationally invariant ground state with $\mathbf{q} = (0,0)$. The spin operator component parallel to the field $S^{\|}$ does not change the number of magnons. The transversal fluctuations created by the perpendicular components $S^{\perp_1}$ and  $S^{\perp_2}$ excite one magnon each. 

 The uniform component of the magnetization
\begin{equation}
\mathbf{M}_{(0,0)} \propto  g \mathbf{S}_{(0,0)}
\end{equation}
for all directions of the fields. The $\mathbf{M}_{(0,0)}$ couples to the $\sf{D_0}$ single magnon excitations. 

The $\sf{D_1}$ couples to the staggered components of the magnetization via $g_s$. When the external field is $H||[001]$, they are given as
\begin{equation}
M^{\perp_1}_{(\pi,\pi)}  \propto  \sum_j (-1)^j S^{\perp_2}_j,
\quad
M^{\perp_2}_{(\pi,\pi)}  \propto  -  \sum_j (-1)^j S^{\perp_1}_j,
\end{equation}
so the $\sf{D_1}$ is present for both Faraday and Voigt configurations.
For the $H||[110]$  and $H||[100]$ field directions they are the same, \begin{equation}
M^{\|}_{(\pi,\pi)} \propto  \sum_j (-1)^j S^{\perp_1}_j,
\quad
M^{\perp_1}_{(\pi,\pi)} \propto -  \sum_j (-1)^j S^{\|}_j.
\end{equation}
and $\sf{D_1}$ is magnetically excited in the Voigt configuration only. 
These selection rules are included into Tab.~\ref{tab:Pmex}.
%\begin{table}[bt]
%\begin{center}
%\begin{ruledtabular}
%\begin{tabular}{cccccc}
%$\mathbf{H}$ & $M^{\|}_{(0,0)}$ & $M^{\|}_{(\pi,\pi)}$ & $\perp$-directions & $M^{\perp}_{(0,0)}$ & $M^{\perp}_{(\pi,\pi)}$   
%\\ 
%\hline
%$H \| [001]$ & ${\sf D_0}$ & & $[100]$ and $[010]$ &  & ${\sf D_1}$ 
%\\ 
%$H \| [110]$ & ${\sf D_0}$ & ${\sf D_1}$ & $[\bar110]$ and $[001]$ & & 
%\\ 
%$H \| [100]$ & ${\sf D_0}$ & ${\sf D_1}$ & $[010]$ and $[001]$ & &  
%\end{tabular}
%\end{ruledtabular}
%\end{center}
%\caption{The same as Tab.~\ref{tab:Pmex}, but for the matrix elements with the oscillating magnetic field of the light. }
%\label{tab:Mmex}
%\end{table}

\subsection{Calculation of the selection rules above saturation from the magnetoelectric coupling}  
%=================================================================

The magnetic component of the light can induce dipolar transitions only. It is the oscillating electric component $\mathbf{E}^\omega$ of the light, coupled to polarization containing higher order spin-quadrupole operators, which can create the two-magnons. To verify this theory, we need to analyze the selection rules in different geometries in detail. 

The two-magnon modes are created by the second order combinations of the perpendicular components. Furthermore, the polarization operators introduced in Eqs.~(\ref{PolOp}) are the sum of the uniform, $\mathbf{q} = (0,0)$, and the staggered, $\mathbf{q} = (\pi,\pi)$ components in the extended Brillouin zone, which can help further to classify excitations. The $\mathbf{q} = (0,0)$  components of the polarizations correspond to the terms proportional to $\cos2\kappa$, and the $\mathbf{q} = (\pi,\pi)$ components are proportional to $\sin2\kappa$ in Eqs.~(\ref{PolOp}). Below we examine the selection rules for the different directions of the external magnetic field.

\subsubsection{The case of $\mathbf{H}\|[001]\|z$}

This is the simplest case, as the parallel spin component is $S^z$, and the perpendicular components $S^{\perp_1}$ and  $S^{\perp_2}$ correspond to $S^x$ and $S^y$. From Eqs.~(\ref{PolOp}) it is clear that only $P^z$ contains second order terms in $S^x$ and $S^y$ -- the $\hat{Q}^{2xy}$ and $\hat{Q}^{x^2 - y^2}$, therefore $P^{z}_{\mathbf{q}} $ excites a two-magnons,  both in the uniform $\mathbf{q} = (0,0)$ and in staggered $\mathbf{q} = (\pi,\pi)$ channels, which we denoted by ${\sf Q}_0$ and ${\sf Q}_1$. 

 From Eqs.~(\ref{PolOp}) it is apparent that the perpendicular components,  $P^x_\mathbf{q} $ and $P^y_\mathbf{q} $ are linear in $S^x$ and $S^y$ for both $\mathbf{q} = (0,0)$ and $\mathbf{q} = (\pi,\pi)$, therefore they can create electrically active magnetic modes. The uniform polarization couples to the ${\sf D_0}$ magnetic mode and the staggered polarization to the ${\sf D_1}$, similar to the case of magnetic transitions.

 \subsubsection{The case of $\mathbf{H}\|[110]\|x$}

The uniform polarization is the largest when the magnetic field is aligned with the $[110]$ crystallographic direction, {\it i.e.} the $x$-axis. Then, the parallel spin component is $S^x$ and the perpendicular ones are $S^y$ and $S^z$. 
%The quadrupolar transitions, therefore, are those which contain any second order combinations of the $y$ and $z$ components. 
Observing Eqs.~(\ref{PolOp}), we see that $\hat{Q}^{2yz}$ appears in ${P}^{x}_{(\pi,\pi)}$ and  ${P}^{y}_{(0,0)}$, and $(S^y)^2$ is present in ${P}^{z}_{(0,0)}$, so the quadrupolar ${\sf Q_0}$ mode can be excited via the perpendicular components of the uniform polarization operator ${P}^{y}_{(0,0)}$ and ${P}^{z}_{(0,0)}$, and the ${\sf Q_1}$ mode is active for the staggered parallel component ${P}^{x}_{(\pi,\pi)}$.
To determine the selection rules for the dipolar transitions, we need to find the components of the polarization operator which are linear in $S^y$ and $S^z$. We get
 that the ${\sf D_0}$ dipolar mode is active for the uniform parallel component ${P}^{x}_{(0,0)}$, while ${\sf D_1}$ can be excited via the staggered perpendicular components ${P}^{y}_{(\pi,\pi)}$ and ${P}^{z}_{(\pi,\pi)}$, as summarized in the second row of Tab.~\ref{tab:Pmex}.

 \subsubsection{The case of $\mathbf{H}\|[100]$}

When the field is set between the $x$ and $y$ axes, along the $[100]$ direction,
we choose the parallel and perpendicular components the spin operators as
\begin{align}
 \left( {S}^{\|}_{j}, {S}^{\perp_1}_{j}, {S}^{\perp_2}_{j} \right)
 =
  \left(\frac{ {S}^x_{j}- {S}^y_{j}}{\sqrt{2}},\frac{ {S}^x_{j}+ {S}^y_{j}}{\sqrt{2}},{S}^z_{j}  \right),
\end{align}
and analogues for the polarization. Thus, the parallel component of the polarization operator becomes
\[
 {P}^{\|}_{(0,0)} 
 \propto \sum_j \hat{Q}^{2\perp_1\perp_2}_{j}\; 
 \quad\text{and}\quad 
 {P}^{\|}_{(\pi,\pi)} \propto \sum_j (-1)^j \hat{Q}^{2\parallel\perp_2}_{j} 
%\overline{ {S}^{\|}_{j}  {S}^{\perp_2}_{j} } 
\;, 
\]
which can create a ${\sf Q_0}$ and a ${\sf D_1}$ excitation (last row of Tab.~\ref{tab:Pmex}), respectively.
%
%We choose $[010]$ as the $\perp_1$, the polarizations are 
%\[
% {P}^{\perp_1}_{(0,0)} \propto \sum_j \hat{Q}^{ 2 \| \perp_2}_j 
% \quad\text{and}\quad 
% {P}^{\perp_1}_{(\pi,\pi)} \propto \sum_j (-1)^j \hat{Q}^{2\perp_1\perp_2}_{j} \;,
%\]
%%
%and they couple to the ${\sf D_0}$ and ${\sf Q_1}$ modes.
%The other perpendicular direction is $[001]$, so $ {P}^{\perp_2}_{\mathbf{q}} =  {P}^{z}_{\mathbf{q}}$ and
%\[
% {P}^{\perp_2}_{(0,0)} \propto  \sum_j \hat{Q}^{ 2 \| \perp_1}_j 
% \quad\text{and}\quad 
% {P}^{\perp_2}_{(\pi,\pi)} \propto  \sum_j (-1)^j  \hat{Q}^{\perp_1^2 - \|^2}_j \;.
%\]
%They excite  ${\sf D_0}$ and ${\sf Q_1}$, just like the ${P}^{\perp_1}$ .
%%
We choose $[010]$ as the $\perp_1$ and $[001]$ as the $\perp_2$ (so $ {P}^{\perp_2}_{\mathbf{q}} =  {P}^{z}_{\mathbf{q}}$), the polarizations are 
\begin{align}
 {P}^{\perp_1}_{(0,0)} &\propto \sum_j \hat{Q}^{ 2 \| \perp_2}_j \;,\\
 {P}^{\perp_2}_{(0,0)} &\propto  \sum_j \hat{Q}^{ 2 \| \perp_1}_j \;,
\end{align}
which couple to the ${\sf D_0}$,  and 
\begin{align}
 {P}^{\perp_1}_{(\pi,\pi)} &\propto \sum_j (-1)^j \hat{Q}^{2\perp_1\perp_2}_{j} \;, \\
 {P}^{\perp_2}_{(\pi,\pi)} &\propto  \sum_j (-1)^j  \hat{Q}^{\perp_1^2 - \|^2}_j \;,
\end{align}
which couple to the ${\sf Q_1}$ modes.

\begin{table*}[bt]
\begin{center}
\begin{ruledtabular}
\begin{tabular}{cccccccccc}
$\mathbf{H}$ & $P^{\|}_{(0,0)}$ & $M^{\|}_{(0,0)}$ & $P^{\|}_{(\pi,\pi)}$ & $M^{\|}_{(\pi,\pi)}$ & $\perp$-directions & $P^{\perp}_{(0,0)}$ & $M^{\perp}_{(0,0)}$ & $P^{\perp}_{(\pi,\pi)}$ & $M^{\perp}_{(\pi,\pi)}$   
\\ 
\hline
$H \| [001]$ & $({\sf Q_0})$ & ${\sf D_0}$ & ${\sf Q_1}$ & & $[100]$ and $[010]$ & ${\sf D_0}$ && ${\sf D_1}$ & ${\sf D_1}$
\\ 
$H \| [110]$ & ${\sf D_0}$ & ${\sf D_0}$ & ${\sf Q_1}$ & ${\sf D_1}$ & $[\bar110]$ and $[001]$ & $({\sf Q_0})$ && ${\sf D_1}$ &
\\ 
$H \| [100]$ & $({\sf Q_0})$ & ${\sf D_0}$ & ${\sf D_1}$ & ${\sf D_1}$ & $[010]$ and $[001]$ & ${\sf D_0}$ && ${\sf Q_1}$ &
\end{tabular}
\end{ruledtabular}
\end{center}
\caption{The branches in the ESR spectrum to which the different components of the polarization and magnetization couple. Since we use unpolarized light, the  perpendicular components of the electric $E_{\perp}^\omega$, which couple to $P^{\perp}$, are  present in both Voigt and Faraday configurations. The electric field $E_{\|}^\omega$, which couples to $P^{\|}$, is present in the Voigt configuration only. Similar considerations hold for the magnetic fields.  An  absorption in the ESR spectrum present in the Voigt configuration, but absent in the Faraday configuration, indicates that it is excited only by the $E_{\|}^\omega$ and/or $H_{\|}^\omega$ component of the light, such as the ${\sf Q_1}$ when $\mathbf{H} \| [001]$ and $\mathbf{H} \| [110]$, and ${\sf D_1}$ when $\mathbf{H} \| [100]$ -- in agreement with the experimental absorption spectra shown in Fig~\ref{fig:ESR846GHz}.  The ${\sf Q_0}$ quadrupolar excitations are silent in the experiment. }
\label{tab:Pmex}
\end{table*}
%-----------------------------------------------------------------------------------------------------------------------

\subsection{Comparing to the experiment}

Table~\ref{tab:Pmex} contains the central theoretical result of our paper: the theoretical calculation based on the magnetoelectric coupling are fully consistent with the experimental observations shown in Fig.~\ref{fig:ESR846GHz}. Namely,
\\
{(i)} For $\mathbf{H}\|[001]$ and $\mathbf{H}\|[110]$, the spin quadrupole operators  in the staggered components of polarizations $\hat{P}^{\|}_{(\pi,\pi)}$ create a ${\sf Q}_1$ excitation with $\mathbf{q}=(\pi,\pi)$ and energy Eq.~(\ref{eq:Q1_x}) leading to the ${\sf Q}_1$ resonance mode in the Voigt, but absent in the Faraday configuration.
\\
{(ii)} For $\mathbf{H}\|[100]$, the spin quadrupole operator in the staggered component in the corresponding $\hat{P}^{\perp}_{(\pi,\pi)}$ creates a ${\sf Q}_1$ excitation visible in both Voigt and Faraday configurations.
\\
{(iii)} The ${\sf D}_1$  mode is absent (very week) in the Faraday configuration when $\mathbf{H}\|[100]$ (topmost line in Fig.~\ref{fig:ESR846GHz}), as it is coupled to the $\hat{P}^{\|}_{(\pi,\pi)}$ and $\hat{M}^{\|}_{(\pi,\pi)}$ only (lowest row in Tab~\ref{tab:Pmex}).
\\
{(iv)} The magnetoelectric effect via the $\hat{P}^{\perp}_{(\pi,\pi)}$ makes the ${\sf D}_1$ mode visible in the Faraday geometry for $\mathbf{H}\|[110]$, as the magnetic coupling is in the $\|$ channel only. 

%Indeed, when $\mathbf{H}\|[110]$ and the spins point along the $x$ direction ($S^{\|}\equiv S^{x}$), the staggered component of the $\hat{P}^{z}_{j}$ %in Eq.~(\ref{eq:polPz}) 
%creates a $(\pi,\pi)$ magnon, giving rise to the ${\sf D}_1$ branch in Fig.~\ref{fig:ESR_100_110}. For $\mathbf{H}\|[100]$, the  spin quadrupole operator in the staggered component of $\hat{P}^{z}_{j}$ creates an $S^{\|}=-1/2$ state with $\mathbf{q}=(\pi,\pi)$ and energy Eq.~(\ref{eq:Q1_x}) leading to the ${\sf Q}_1$ resonance mode, in full agreement with the matrix elements shown in Table\ref{tab:Pmex}.

Let us also mention that the ${\sf D}_1$ magnon mode is usually not observed in the ESR spectra, unless Dzyaloshinskii-Moriya interaction or staggered $g$-tensor is present \cite{ZvyaginPRL2014}. In SCGO, not only the staggered $g$-tensor, but also the staggered component of the polarization can lead to absorption at $\mathbf{q}=(\pi,\pi)$.

The uniform component of the polarizations can  create quadrupole excitations at $\mathbf{q}=(0,0)$ as well, which we denoted as ${\sf Q_0}$. However, this mode is not seen in the experiment. The on-site energy of the $\Delta S^z=-2$ state is deep inside the continuum at $\mathbf{q}=(0,0)$, and due to interactions with the two-magnon continuum, it decays very quickly, making it unobservable (only for very large anisotropy, $\Lambda/J \agt 7.5$, does the quadrupolar mode ${\sf Q_0}$ split off from the continuum). 
 This is in sharp contrast to the case of the ${\sf Q_1}$, when the continuum  shrinks to a single point at $\mathbf{q}=(\pi,\pi)$ with an energy $\omega_{{\sf D}_0}+\omega_{{\sf D}_1}$ different from $\omega_{{\sf Q}_1}$. In addition, the ${\sf Q_1}$ state at $\mathbf{q}=(\pi,\pi)$ is fully decoupled from the two-magnon continuum \cite{Silberglitt1970}, resulting in a sharp absorption peak at $\omega_{{\sf Q}_1}$. 

Although multipole fluctuations have long been theoretically proposed~\cite{quad_theo1,quad_theo2,quad_theo3,quad_theo4}, the ${\sf Q}_1$ mode in the saturated state of SCGO is, up to our knowledge, the first unambiguous experimental observation of a purely quadrupole excitation in a quantum magnet.

%%%%%%%%%%%%%%%%%%%%%%%%%%%%%%%%%%%%%%%%%%%%%%%%
\section{Conclusion}
\label{sec:conlusion}
%%%%%%%%%%%%%%%%%%%%%%%%%%%%%%%%%%%%%%%%%%%%%%%%

In conclusion, we studied multipole excitations in  single crystals of the magnetoelectric insulator, SCGO. The observation of two-magnons is not new, they have been first observed in FeI$_2$ \cite{FeI2}, in the  spin-1 chain compound, NiCl$_2$-$4$SC(NH$_2$)$_2$~\cite{Zvyagin_DTN}, as well as in the excitation spectrum of the other \aa kermanites~\cite{PencPRL} and even in ultracold atomic systems \cite{2magnoncoldatom}.  However, in those materials the single-ion anisotropy is the dominant term, reducing the spin length considerably and mixing the dipolar and quadrupolar degrees of freedom. As a result, magnon and two-magnon modes have both dipolar and quadrupolar character, and couple to both magnetic and electric components of the exciting light in earlier experiments.  
 
The nearly isotropic property of SCGO, on the other hand, allows for the emergence of uniquely pure quadrupolar excitations, appearing completely detached from magnetic transitions.
Using multifrequency ESR technique, supported by theoretical investigation of transition matrix elements, we clarified the quadrupolar nature of the two-magnon excitation. %
Furthermore, based on analytical description of the excitations, available in the high field regime, and utilizing the measured saturation fields, we extracted the magnetic coupling  values, as well as the components of the $g$-tensor with high precision.
The spin wave spectra with the extracted parameter values are in excellent agreement with the measurements, throughout the entire magnetic field regime and for each field direction.

It is the fortunate constellation of several properties found in Sr$_2$CoGe$_2$O$_7$ that made the observation of the quadrupolar waves possible: (i) The non-centrosymmetric position of the  Co$^{2+}$ ions allowing for the  magnetoelectric coupling; (ii) This coupling is realized on single-sites, involving quadratic spin operators, which are finite due to the large $S=3/2$ spin of Co$^{2+}$ ion, and can simultaneously excite two magnons on one site; (iii) The two magnons form a bound state at the momentum $\mathbf{q}=(\pi,\pi)$ in the Brillouin zone, well separated from the two-magnon continuum to prevent its decay and to make it apeear as a single mode; (iv) The alternating local environment of the magnetic ions results in a finite staggered polarization necessary to create this bound state of two magnons, the two-magnon excitation ${\sf Q}_1$ we observed; (v) Small anisotropy to ensure the separation of the spin dipolar and quadrupolar degrees of freedom.

Our investigations provide a basis for further studies of emerging multipolar excitations with the use of well-spread experimental approaches, such as the ESR in the present work. 
Detecting such modes could help designing new magnetoelectric devices, in which electrically active magnetic excitations may carry and store information.

Furthermore, it may serve as a guide in the quest for nematic and more exotic, otherwise  `hidden' orders. By observing static properties such phases are usually experimentally undetectable due to the lack of magnetic ordering. An alternative route to reveal nematic phases is  probing the dynamical properties and looking for the signatures of condensation of quadrupolar excitations at high magnetic fields, a prerequisite for the formation of spin nematic phase~\cite{nematic_condensation,Zhitomirsky2010,J1J2chainstheory,Lauchli2006, SatoPRL}. For example, in LiCuVO$_4$, a high field phase just below the saturation is believed to be a spin nematic phase~\cite{Svistov}. The NMR spin relaxation measurements showed a decay coming from excitations with twice as large slope as conventional magnons above the saturation field, indirectly supporting the condensation of such nematic waves~\cite{Buttgen}. The nonmagnetic nature of those excitation was recently shown by another NMR experiment~\cite{Orlova}, however without explicitly proving the breaking of the O(2) symmetry, a defining property of the  nematic state.

%%%%%%%%%%%%%%%
\begin{acknowledgments}
We thank K. Yamauchi, T. Oguchi, and M. Zhitomirsky for helpful discussions.
This work was supported in part by the Grants-in-Aid for Scientific Research (Grant Nos. JP15K05145, JP25220803, JP24244059, and JP25246006) from MEXT, Japan and by JSPS Core-to-Core Program, A Advanced Research Networks and by the Hungarian OTKA Grant No. K106047. K.P. gratefully acknowledges support from the International Joint Research Promotion Program (Short Stay) of the Osaka University.
\end{acknowledgments}
%%%%%%%%%%%%%%%

\appendix

%%%%%%%%%%%%%%%
\section{The multiboson spin wave}\label{app:multib}
%%%%%%%%%%%%%%
 In the next part we outline a simplified multiboson approach, in which we take a small single-ion anisotropy ($ \Lambda $) limit. By comparing our analytical solution to the measured spectrum we further prove the isotropic nature of this material.

%================================================================ 
\subsection{Variational setup}

Here we will consider the Hamiltonian (\ref{eq:Hamiltonian}) assuming $J_{pz}=0$ for simplicity. We closely follow the derivation presented in Ref.~\cite{juditPRB}. 
%We consider the following Hamiltonian,
%\begin{align}
%\mathcal{H} =& 
%  J \sum_{\langle i,j \rangle} \mathbf{S}_i \cdot \mathbf{S}_j +  \Lambda \sum_{i} (S_i^z)^2\nonumber\\
%&- \mu_{\rm B} \sum_{i} (g_{ab} H^x S_i^x+g_{z} H^z S_i^z) \;.
%\label{eq:HamiltonianX}
%\end{align}
Based on numerical calculations we need two variational parameters to characterize the ground state wave function, and consequently the bosons representing the excited states, for all values of magnetic field. 

We rotate the usual $\left|m^z\right>$ basis, with $m^z=\left<S^z\right>=-\frac 3 2,\hdots,\frac 3 2$, to a new basis, in which one of the states corresponds to the ground state. We perform such rotation on both sub-lattices in the following way
\begin{subequations}
\begin{eqnarray}
&&\textstyle{\!\left|\psi_0\right>_A\!\!=\!\frac{\left|\frac{3}{2}\right>\!-\!i \sqrt{3}\eta \left|\frac{1}{2}\right>\!-\!\sqrt{3}\eta \left|\frac{\!-1\!}{2}\right>\!+\!i\left|\frac{\!-3\!}{2}\right>}{\sqrt{6 \eta^2+2}} }\;,\\
&&\textstyle{\!\left|\psi_1
\right>_A\!\!=\!\frac{-i \sqrt{3}\eta\left|\frac{3}{2}\right> - (2\eta\!-\!1) \left|\frac{1}{2}\right> -i (2\eta\!-\!1) \left|\frac{\!-1\!}{2}\right> - \sqrt{3}\eta \left|\frac{\!-3\!}{2}\right>}{\sqrt{14\eta^2-8\eta+2}} }\;,\\
&&\textstyle{\!\left|\psi_2\right>_A\!\!=\!\frac{-\sqrt{3}\eta \left|\frac{3}{2}\right>\!-\!i  \left|\frac{1}{2}\right>\!-\! \left|\frac{\!-1\!}{2}\right>\!-\!i\sqrt{3}\eta\left|\frac{\!-3\!}{2}\right>}{\sqrt{6 \eta^2+2}} }\;,\\
&&\textstyle{\!\left|\psi_3
\right>_A\!\!=\!\frac{i(2\eta\!-\!1) \left|\frac{3}{2}\right> - \sqrt{3}\eta \left|\frac{1}{2}\right> -i  \sqrt{3}\eta \left|\frac{\!-1\!}{2}\right> +(2\eta\!-\!1) \left|\frac{\!-3\!}{2}\right>}{\sqrt{14\eta^2-8\eta+2}} }\;,
\end{eqnarray}\label{eq:trafo_A}
\end{subequations}
The transformation on sub-lattice $B$ corresponds to the complex conjugate of Eqs.~(\ref{eq:trafo_A}). For $\eta=1$ these states represent the $m^y=\frac 3 2,\hdots,\frac 3 2$ basis, with $\left|\psi_0\right>_A=\left|m^y\right>=-\frac 3 2$ and $\left|\psi_0\right>_B=\left|m^y\right>=\frac 3 2$. In other words, we changed the quantization axis to $y$. The value of $\eta$ affects the length of the spin, which can be expressed as $S=\frac{3\eta(1+\eta)}{1+3\eta^2}$. In fact, below the saturation, due to the single-ion anisotropy term, $\eta$ differs from 1 the spin length is shorter than  $\frac 3 2$. 

We consider two cases, when the field is in the $xy$-plane and when it is perpendicular to it. In the first case we set the field along the $x$-axis, that is along the $[110]$ crystallographic direction. As long as no $(P^z_i\!\cdot\!P^z_j)$ terms are considered, all in-plane directions are equivalent, and the spectrum looks the same --as far as the energy levels are concerned-- for the directions $[110]$ and $[100]$. 

To include the effect of magnetic field on the ground state, we need to allow the spins to turn  away from the $y$ axis, towards the direction of magnetic field. 

In case of $H\|[110]$, therefore, we apply an additional rotation about the $z$-axis with the angles $\pm\vartheta$ on sub-lattice $A/B$. The full variational setup for $H\|[110]$ then has the form of $\left|\Psi_i\right>_A=e^{-i \vartheta S^z}\left|\psi_i\right>_A$, and $\left|\Psi_i\right>_B=e^{i \vartheta S^z}\left|\psi_i\right>_B$, ($i=0,\hdots,3$).

For perpendicular field, $H\|[001]$, we need a rotation about the $x$-axis with the angles $\mp\vartheta$ on sub-lattice $A/B$. The variational setup for $H\|[001]$ becomes $\left|\Psi_i\right>_A=e^{i \vartheta S^x}\left|\psi_i\right>_A$, and $\left|\Psi_i\right>_B=e^{-i \vartheta S^x}\left|\psi_i\right>_B$, ($i=0,\hdots,3$). 

For both field directions the angle $\vartheta$ changes from $0$ to $\pi/2$ as the spins tilt from the original $y$-axes towards the corresponding field direction.
Evidently, when the spins are saturated along the $x(z)$ direction $\eta$ becomes 1 and $\vartheta=\pi/2$. In these simple cases, the new basis corresponds to the states $\left<S^{x(z)}\right>=-\frac 3 2,\hdots,\frac 3 2$ on both sub-lattices, the ground state becomes $\prod_{u.c.}\!\left|\Psi_0\right>_A\!\left|\Psi_0\right>_B=\prod_{u.c.}\left|m^{x(z)}=\frac 3 2\right>_A\left|m^{x(z)}=\frac 3 2\right>_B$, and creating a state $\left|\Psi_n\right>$ with $n=1,2$ or $3$ on any of the sub-lattices corresponds to  a transition of $\Delta S^{x(z)}=n=1,2$ or $3$, therefore, a dipolar, quadrupolar or octupolar transition.

We determine the variational parameters $\eta$ and $\vartheta$ by minimizing the ground state energy $\left<{\sf GS}\right| \mathcal{H}\left|{\sf GS}\right>$ with $\left|{\sf GS}\right>= \prod_{u.c.}\!\left|\Psi_0\right>_A\left|\Psi_0\right>_B$.

%==================================================================
\subsection{Static properties}

To determine the magnetization and induced polarization as a function of external magnetic field, we calculate the expectation values of the spin and polarization operators~\ref{PolOp} in the ground state. For in-plane magnetic field, the ground state is $\left|\Psi_0\right>_A=e^{-i \vartheta S^z}\left|\psi_0\right>_A$, and $\left|\Psi_0\right>_B=e^{i \vartheta S^z}\left|\psi_0\right>_B$ only $P^z$ is finite, and takes the value:

\begin{equation}
P^z=\frac{6 \eta\cos(2 (\kappa-\varphi))}{1+3\eta^2}\;.\label{eq:Pz_exp}
\end{equation}
Substituting the values for the variational parameters $\eta$ from Eq.~\ref{eq:eta} and $\vartheta$ from Eq.~\ref{eq:vartheta} we plotted the theoretical induced polarization. Of course, including the polarization--polarization term would give a better agreement with the experimentally measured polarization and would reproduce the drop of $P^z$ to zero at small field.~\cite{RomPRB}

%%%%%%%%%%%%%   FIG     %%%%%%%%%%%%%%%%%%%%%%
\begin{figure}[h!]
\begin{center}
\includegraphics[width=0.8\columnwidth]{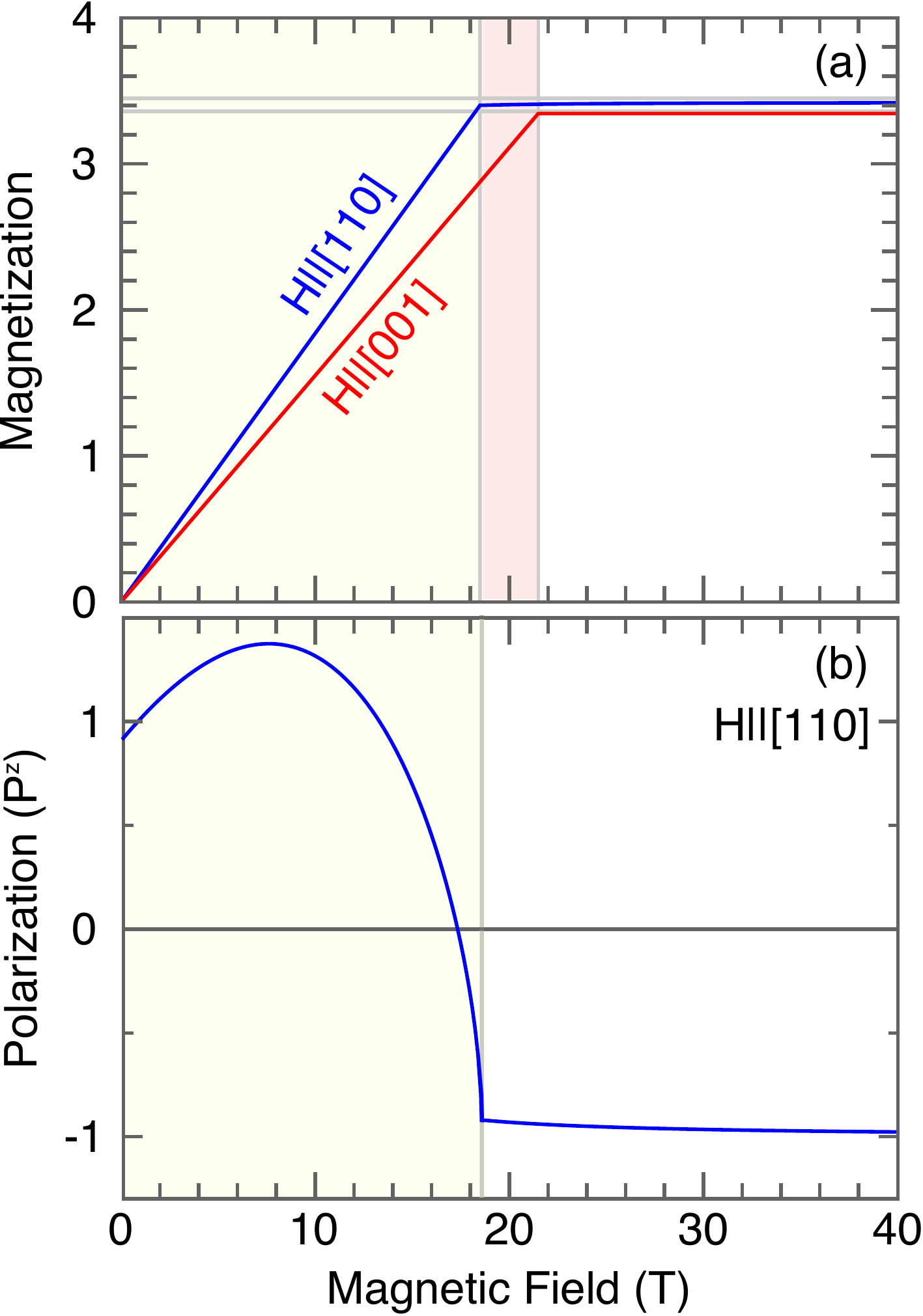}
\caption{(a) Calculated magnetization using our variational setup using the variational parameters determined below in Sec.~\ref{sec:in_plane} and Sec.~\ref{sec:outof_plane} for in-plane and out-of-plane magnetic fields, respectively. (b) Theoretically calculated induced polarization in the simplified picture using Eq.~\ref{eq:Pz_exp}. We neglected the polarization-polarization term in the Hamiltonian, and use the small $ \Lambda $ expansion. If $P^z_i\cdot P^z_j$ were kept, the polarization would drop to zero at zero field. For both figures we used the $g$  and interaction values from the linear fit in Eqs.~\ref{eq:gs} and~\ref{eq:JandLambda} in the main text.}
\label{fig:pz}
\end{center}
\end{figure}
%%%%%%%%%%%%%%%%%%%%%%%%%%%%%%%%%%%%%%%

%==================================================================
\subsection{In-plane magnetic field}\label{sec:in_plane}

%%%%%%%%%%%%%   FIG     %%%%%%%%%%%%%%%%%%%%%%
\begin{figure}[h!]
\begin{center}
\includegraphics[width=0.9\columnwidth]{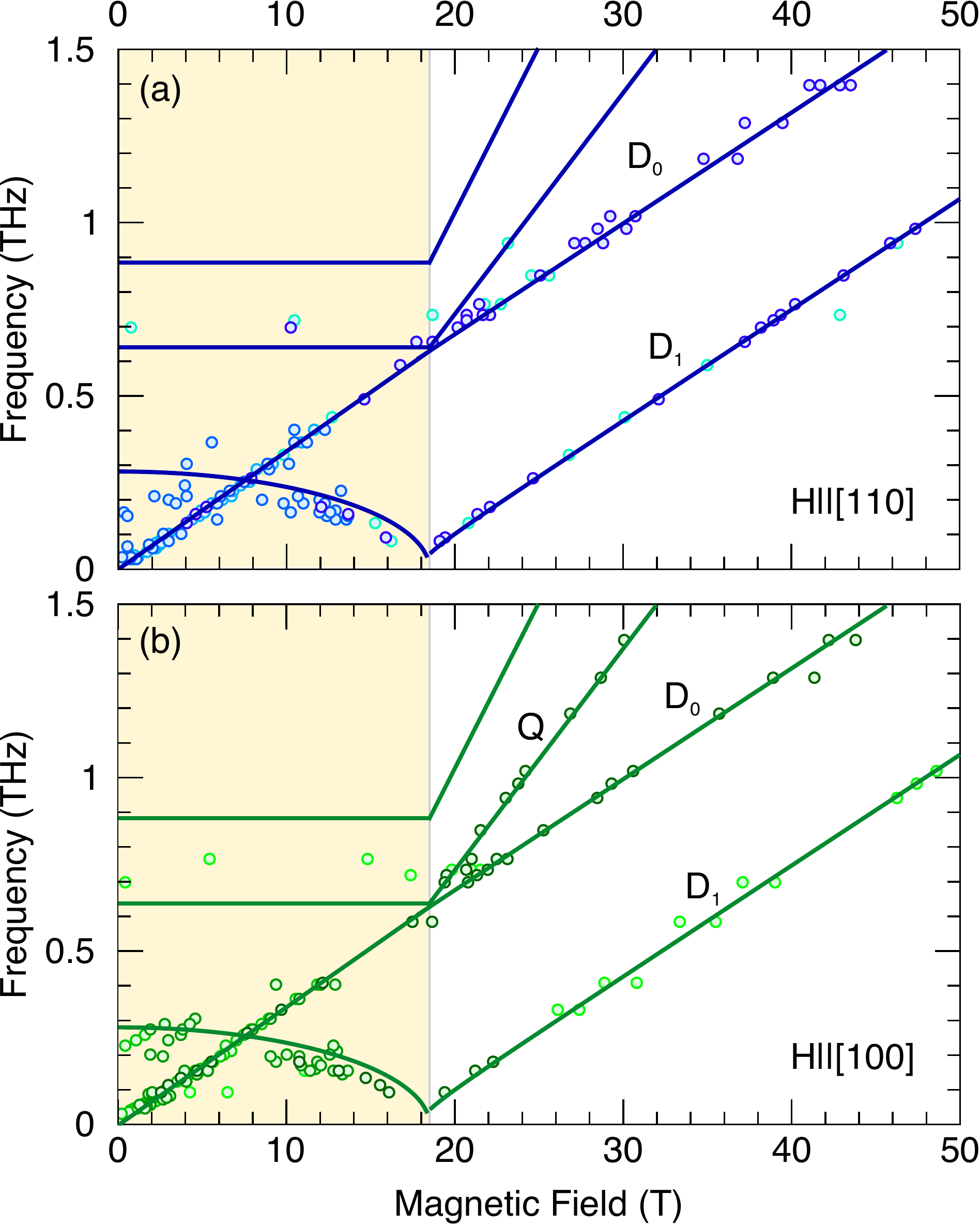}
\caption{Simplified multiboson approach compared with the measured ESR spectrum for field directions parallel to the $H\|[110]$ and $H\|[100]$ crystallographic direction in (a) and (b), respectively. The energy of dipole and quadrupole excitation is in very good agreement with the experimental findings.
}
\label{fig:h_ab}
\end{center}
\end{figure}
%%%%%%%%%%%%%%%%%%%%%%%%%%%%%%%%%%%%%%%

For fields lying in the $xy$-plane the ground state energy has the following form 
\begin{eqnarray}
E_0&=&-36 J\cos2\vartheta\frac{(1+\eta)^2\eta^2}{(1+3\eta^2)^2}
-6 g_{ab} H^x\sin\vartheta\frac{\eta(1+\eta)}{1+3\eta^2}\nonumber\\
&&+ \Lambda  \frac{3(3+\eta^2)}{2(1+3\eta^2)}\;.
\end{eqnarray}

Although this can be minimized analytically, we restrict ourselves to the limit, in which $ \Lambda $ is a small parameter. %To prove the validity of this limit, we plot the result of a numerical minimization together with our closed formulae.
The variational parameters take the following form:
\begin{eqnarray}
\vartheta=\left\{\begin{aligned}
 & \textstyle{\arcsin\frac{g_{ab} H^x}{12J}}\;,\qquad &g_{ab}H^x< 12 J\\
&\pi/2\;, \qquad &g_{ab} H^x\ge 12 J
\end{aligned}
\right.\nonumber\\
\label{eq:vartheta}
\end{eqnarray}
and
\begin{eqnarray}
\eta=\left\{\begin{aligned}
 &\textstyle{1+\frac{ \Lambda }{6 J}}\;,\qquad &g_{ab}H^x< 12 J\\
&\textstyle{1+\frac{ \Lambda }{g_{ab} H^x-6J}}\;, \qquad &g_{ab}H^x\ge 12 J
\end{aligned}
\right.\nonumber\\
\label{eq:eta}
\end{eqnarray}
At zero field the spins are antiparallel along the $y$-axis. In finite field the spins start to tilt towards each-other to align themselves with the field direction. $H^x_c=12 J/g_{ab}$ gives the transition field, at which the spins become parallel to each-other and $H^x$ as well. Above  $H^x_c$ their length grows asymptotically approaching the maximal $3/2$ value. 

Taking the small $ \Lambda $ limit, we can continue our analysis with the excitation spectrum, in which the different multipole excitations decouple from each-other and we get simple equations of motion describing them separately. 

As the next step, we express each operator in the Hamiltonian~\ref{eq:Hamiltonian} in terms of our new basis, using the transformation $\hat U=\left(\Psi_0,\Psi_1,\Psi_2,\Psi_3\right)$. in which the columns correspond to the component vectors of the new states $\left|\Psi_i\right>$, $i=0,\hdots,3$.

We introduce a boson for each of the states $a^\dagger_{i,L}\left|0\right>=\left|\Psi_i\right>_L$, where $i=0,\hdots,3$ and $L=\{A,B\}$. The mean-field ground state is the product state  $\prod_{u.c.}\left|\Psi_0\right>_A\left|\Psi_0\right>_B$ in which the boson  $a^\dagger_{0,L}$ is condensed on both sub-lattices. Creating a boson  $a^\dagger_{n,L}$ with $n=1,2$ or $3$ corresponds to an excitation with dipole, quadrupole or octupole character, respectively. 

Let us start with the dipole excitation. The spin-wave Hamiltonian for these modes has the form

\begin{eqnarray}
\mathcal{H}_{\sf D}=
\!\!\left(
\begin{array}{c}
a^\dagger_{1,A}\\
a^\dagger_{1,B}\\
a^{\phantom{\dagger}}_{1,A}\\
a^{\phantom{\dagger}}_{1,B}
\end{array}
\right)^T\!\!
\!\left(
\begin{array}{cccc}
\varepsilon_{0} & f_{+} & 0 & f_{-} \\ 
 f_{+}  & \varepsilon_{0} & f_{-} & 0\\
0 & f_{-} & \varepsilon_{0} &   f_{+} \\
f_{-}& 0 &  f_{+}  & \varepsilon_{0}
\end{array}
\right)\!
\!\!\left(
\begin{array}{c}
a^{\phantom{\dagger}}_{1,A}\\
a^{\phantom{\dagger}}_{1,B}\\
a^\dagger_{1,A}\\
a^\dagger_{1,B}
\end{array}
\right)\!\!,
\label{eq:b_Ham}
\end{eqnarray}
where $\varepsilon_{0}=6 J \cos 2 \vartheta+g_{ab} H_x\sin\vartheta+ \Lambda $ and $ f_{\pm} =3J\eta\cos(2\vartheta) \pm 3J_z(\eta-2)$.
Inserting the solutions Eqs.~\ref{eq:vartheta} and ~\ref{eq:eta} we get
%\begin{eqnarray}
%\varepsilon=\left\{\begin{aligned}
% &6J+D\;,\qquad &H^x< 12 J/g_{ab}\\
%&-6J+D+g_{ab} H^x\;, \qquad &H^x\ge 12 J/g_{ab}
%\end{aligned}
%\right.\nonumber\\
%\label{eq:ed}
%\end{eqnarray}
%
%The dipole excitation energies can be expressed as 
%\begin{eqnarray}
%\omega_{\sf D_{0/1}}=\sqrt{(\varepsilon_{\vartheta}\mp6J(\eta-2))(\varepsilon_{\vartheta}\mp6J\eta\cos2\vartheta)}
%\label{eq:omega_d}
%\end{eqnarray}
%%
%More explicitly, 
%%

\begin{widetext}
\begin{eqnarray}
\omega_{\sf D_{0}}=\left\{\begin{aligned}
 &\textstyle{g_{ab} H^x\sqrt{1+\frac{ \Lambda }{6J}}\sqrt{1+\frac{J_z-J}{2J}\left(1-\frac{\Lambda}{6 J}\right)}}\;,\qquad &g_{ab}H^x< 12 J\\
&\textstyle{g_{ab}H^x -3J+3J_z+ \Lambda\left(\frac{J+J_z}{2J_z}-\frac{3(J-J_z)}{g_{ab}H^x}\right) }\;, \qquad & g_{ab}H^x\ge 12 J
\end{aligned}
\right.\nonumber\\
\label{eq:D0_mode}
\end{eqnarray}
and
\begin{eqnarray}
\omega_{\sf D_{1}}=\left\{\begin{aligned}
 &\textstyle{(6J+\Lambda)\sqrt{2\left(1-\left(\frac{g_{ab}H^x}{12J}\right)^2\right)\left(1+\frac{J_z}{J}-\frac{12 J_z}{6J+\lambda}\right)}}\;,\quad & g_{ab} H^x< 12 J\\
& \textstyle{g_{ab} H^x-9J-3J_z+ \Lambda\left(1-\frac{(J-J_z)(g_{ab}H^x-6J_z)}{2J_z(g_{ab}H^x-12J_z)}\right) }\;, \quad & g_{ab} H^x\ge 12 J
\end{aligned}
\right.\nonumber\\
\label{eq:D1_mode}
\end{eqnarray}
\end{widetext}
The quadrupolar and octupolar modes are even more straightforward, as the Hamiltonian describing their dynamics is diagonal and we can directly read off the energies.

The two quadrupole modes, $a^\dagger_{2,A}$ and $a^\dagger_{2,B}$, are degenerate and have the energies 
\begin{equation}
\omega_{\sf Q}=12 J \cos 2 \vartheta+2 g_{ab} H_x\sin{\vartheta}+\frac{ \Lambda }{2}(3\eta-1)
\label{eq:omega_q}
\end{equation}
Substituting the solutions Eqs.~\ref{eq:vartheta} and~\ref{eq:eta} we get

\begin{eqnarray}
\omega_{\sf Q}=\left\{\begin{aligned}
 &12 J+ \Lambda \;,\quad & g_{ab}H^x< 12 J\\
&2 g_{ab}H^x-12 J+ \Lambda \;, \quad & g_{ab}H^x\ge 12 J
\end{aligned}
\right.\nonumber\\
\label{eq:Q_mode}
\end{eqnarray}

The octupole modes $a^\dagger_{3,A}$ and $a^\dagger_{3,B}$ are also degenerate with the energy
%
%\begin{equation}
%\omega_{\sf O}=18 J \cos 2 \vartheta+3 g_{ab} H_x\sin{\vartheta}+\frac{3  \Lambda }{2}(\eta-1)
%\label{eq:omega_o}
%\end{equation}
%Which after using the solutions for $\vartheta$ and $\eta$ becomes
\begin{eqnarray}
\omega_{\sf O}=\left\{\begin{aligned}
 &18 J\;,\qquad & g_{ab}H^x< 12 J\\
&3 g_{ab}H^x-18 J\;, \qquad & g_{ab}H^x\ge 12 J
\end{aligned}
\right.\nonumber\\
\label{eq:O_mode}
\end{eqnarray}
%%%%%%%%%%%%%%   FIG     %%%%%%%%%%%%%%%%%%%%%%
%\begin{figure}[h!]
%\begin{center}
%\includegraphics[width=0.8\columnwidth]{H110_all.eps}
%\caption{Simplified multiboson approach compared with the measured spectrum for field parallel to the $x$-axis ($H\|[110]$).}
%\label{fig:h110}
%\end{center}
%\end{figure}
%%%%%%%%%%%%%%%%%%%%%%%%%%%%%%%%%%%%%%%%

In Fig.
~\ref{fig:h_ab} we plot the measured spectrum and the analytical solutions~\ref{eq:D0_mode},~\ref{eq:D1_mode},~\ref{eq:Q_mode} and~\ref{eq:O_mode} using the model parameters $g_{ab}=2.28$ and Eqs.~\ref{eq:JandLambda}  determined from experimental data as explained in the main text.
We find an excellent agreement between our simple model and the experiment, further justifying our assumption for the large field limit and the determined parameters.

%==================================================================
\subsection{Perpendicular field}\label{sec:outof_plane}

%%%%%%%%%%%%%   FIG     %%%%%%%%%%%%%%%%%%%%%%
\begin{figure}[h!]
\begin{center}
\includegraphics[width=0.95\columnwidth]{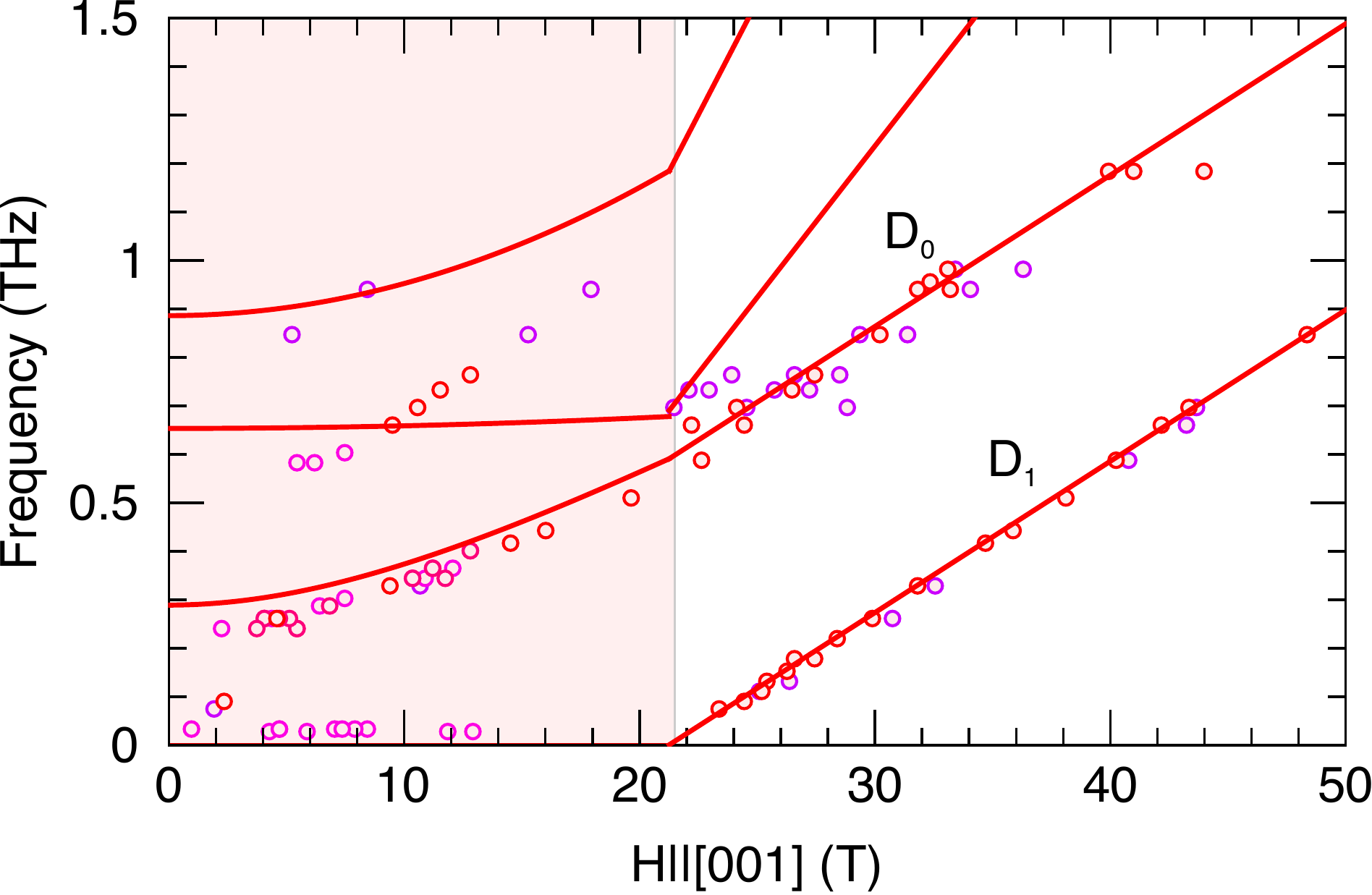}
\caption{Simplified multiboson approach compared with the measured spectrum for field parallel to the $H\|[001]$ crystallographic direction.}
\label{fig:h001}
\end{center}
\end{figure}
%%%%%%%%%%%%%%%%%%%%%%%%%%%%%%%%%%%%%%% 

For magnetic field applied perpendicular to the layers, the angle $\vartheta$ measures the canting from the $y$-axis towards the $z$-axis and the ground state energy becomes 

\begin{eqnarray}
E_0&=&-36 \frac{(1+\eta)^2\eta^2}{(1+3\eta^2)^2}(J\cos^2\vartheta-J_z\sin^2\vartheta)\nonumber\\
&&+ \Lambda  \frac{3(3+\eta^2)}{2(1+3\eta^2)}+ \Lambda \frac{3(1+\eta)(3\eta-1)}{(1+3\eta^2)}\sin^2\vartheta\nonumber\\
&&-6 g_{z} H^z\sin\vartheta\frac{\eta(1+\eta)}{1+3\eta^2}\;.
\end{eqnarray}
We again look for the solution up to leading order in $ \Lambda $, and obtain
\begin{eqnarray}
\vartheta=\left\{\begin{aligned}
 & \arcsin\delta\;,\qquad &\delta< 1\\
&\pi/2\;, \qquad &\delta\ge 1
\end{aligned}
\right.\nonumber\\
\label{eq:vartheta_z}
\end{eqnarray}
where we introduced the parameter $\delta=\frac{g_{z} H^z}{6J+6J_z+2 \Lambda }$ which gives the transition field upon becoming unity. For $\delta=1$ the magnetic field reaches the value $H^z= \frac{6J+6J_z+2 \Lambda }{g_{z}}$ at which the spins become parallel to each-other as well as the magnetic field.  
\begin{eqnarray}
\eta=\left\{\begin{aligned}
 &\textstyle{1+\frac{ \Lambda }{6 J}(1-\delta^2)}\;,\qquad & \delta< 1\\
&1\;, \qquad & \delta\ge 1
\end{aligned}
\right.\nonumber\\
\label{eq:eta_z}
\end{eqnarray}

The spin-wave Hamiltonian describing the dynamics of dipole excitations can be written in the same form as Eq.~\ref{eq:b_Ham} with the parameters $\varepsilon_{0}=-6J_z+6(J+J_z)\cos^2\vartheta+\frac{\Lambda}{2}(3\cos(2\vartheta)-1)+g_c H^z\sin\vartheta$, $f_{\pm}=3J\eta\pm3(J-J_z)\left(1-\frac{\eta}{2}\right)\mp3(J+J_z)\cos(2\vartheta)\left(1-\frac{\eta}{2}\right)$.

%The energies for the dipole excitations can be expressed as
%\begin{eqnarray}
%\omega_{\sf D_{\pm}}\!=\!\sqrt{(\varepsilon_{\eta,\vartheta}\!\pm\!(f_{\eta,\vartheta}\!+\!g_{\eta,\vartheta}))(\varepsilon_{\eta,\vartheta}\!\pm\!(g_{\eta,\vartheta}\!-\!f_{\eta,\vartheta}))}
%\label{eq:omega_d_Z}
%\end{eqnarray}
%
%Inserting Eqs.~(\ref{eq:vartheta_z}) and~(\ref{eq:eta_z}) we get
%\begin{eqnarray}
%\varepsilon=\left\{\begin{aligned}
% &6J\eta\;,\qquad &\delta< 1\\
%&-6J-2D+g_{z} H^z\;, \qquad &\delta\ge 1
%\end{aligned}
%\right.\nonumber\\
%\label{eq:ed}
%\end{eqnarray}
%
%\begin{eqnarray}
%f+g=\left\{\begin{aligned}
% &-6J\eta\;,\quad &\delta<1\\
%&-6 J\;, \quad &\delta\ge 1\;,
%\end{aligned}
%\right.\nonumber\\
%\label{eq:f+g}
%\end{eqnarray}
%and
%\begin{eqnarray}
%g-f=\left\{\begin{aligned}
% &-6J(\eta-2)(1-2\delta^2)\;,\quad &\delta<1\\
%&-6 J\;, \quad &\delta\ge 1\;.
%\end{aligned}
%\right.\nonumber\\
%\label{eq:g-f}
%\end{eqnarray}
%It is apparent that below saturation $\omega_+$ becomes zero, as  $\varepsilon=-f+g$, as written in Eqs.~(\ref{eq:ed}) and~(\ref{eq:f+g}). 
The energies of the dipolar excitations can be easily calculated using Eqs.~\ref{eq:vartheta_z}-~\ref{eq:eta_z}.
\begin{eqnarray}
\omega_{\sf D_1}=\left\{\begin{aligned}
 &0\;,\qquad & \delta< 1\\
&-6J-6Jz+g_cH^z-2 \Lambda \;, \qquad & \delta\ge 1
\end{aligned}
\right.\nonumber\\
\label{eq:D1_modeZ}
\end{eqnarray}
and
\begin{widetext}
\begin{eqnarray}
\omega_{\sf D_0}=\left\{\begin{aligned}
 &\textstyle{\sqrt{6J+\Lambda(1-\delta^2)}\sqrt{24J-12(J+J_z)(1-\delta^2)(1-\frac{\Lambda(1-\delta^2)}{6J})}}\;,\qquad & \delta< 1\\
&6J-6J_z+g_c H^z-2  \Lambda \;, \qquad & \delta\ge 1
\end{aligned}
\right.\nonumber\\
\label{eq:D0_modeZ}
\end{eqnarray}
where $\eta$ is the solution below the transition field ($\delta<1$).  

\end{widetext}

The spin-wave Hamiltonian for quadrupole and octupole excitations become diagonal for $H\|[001]$ too, and we can easily read off the eigenvalues. Similarly to the in-plane field case, the modes are two-fold degenerate.
\begin{eqnarray}
\omega_{\sf Q}=\left\{\begin{aligned}
 &\textstyle{12 J+ \Lambda +\frac{ \Lambda ^2}{4 J}+ \Lambda \delta^2(1-\frac{ \Lambda }{2J})}\;,\qquad & \delta< 1\\
&2 g_{z}H^z-12 J_z-2 \Lambda \;, \qquad & \delta\ge 1
\end{aligned}
\right.\nonumber\\
\label{eq:Q_modeZ}
\end{eqnarray}

\begin{eqnarray}
\omega_{\sf O}=\left\{\begin{aligned}
 &\textstyle{18 J+\frac{ \Lambda ^2}{4 J}+6 \Lambda \delta^2(1-\frac{ \Lambda }{12J})}\;,\qquad & \delta< 1\\
&3 g_{z}H^z-18 J_z\;, \qquad & \delta\ge 1
\end{aligned}
\right.\nonumber\\
\label{eq:O_modeZ}
\end{eqnarray}

In Fig.~\ref{fig:h001} we plot the calculated energies together with the measured spectrum for $H\|[001]$ and find excellent agreement between our simple model and the observed spectrum.

%\end{appendix}

%%%%%%%%%%%%%%%


\begin{thebibliography}{99}
\bibitem{conversion}
We use  1~THz=48.0~K and $\mu_{\rm B} = 0.6717$~K~T$^{-1}$=13.996~GHz~T$^{-1}$ for unit conversion.
\end{thebibliography}
\end{document}